\documentclass{lmcs} 
\usepackage[utf8]{inputenc}
\pdfoutput=1

\usepackage{lastpage}
\lmcsdoi{19}{3}{8}
\lmcsheading{}{\pageref{LastPage}}{}{}%
{Oct.~14,~2021}{Jul.~26,~2023}{}


\usepackage{hyperref}
\theoremstyle{plain} 

\usetikzlibrary{patterns}

\usepackage{algorithm2e}

\usepackage{mathrsfs}

\DeclareSymbolFont{symbolsC}{U}{txsyc}{m}{n}
\SetSymbolFont{symbolsC}{bold}{U}{txsyc}{bx}{n}
\DeclareFontSubstitution{U}{txsyc}{m}{n}
\DeclareMathSymbol{\multimapdotboth}{\mathrel}{symbolsC}{22}

\usepackage{stmaryrd}

\usetikzlibrary{shapes.misc}

\begin{document}

\title[A Spatial Logic for Simplicial Models]{A Spatial Logic for Simplicial Models}

\author[M.~Loreti]{Michele Loreti\lmcsorcid{0000-0003-3061-863X}\rsuper{a}}	
\address{University of Camerino, School of Science and Technology, Camerino, Italy}	
\email{michele.loreti@unicam.it}  

\author[M.~Quadrini]{Michela Quadrini\lmcsorcid{0000-0003-0539-0290}\rsuper{b}}	
\address{University of Camerino, School of Science and Technology, Camerino, Italy}	
\email{michela.quadrini@unicam.it}  





\begin{abstract}
  \noindent Collective Adaptive Systems often consist of many heterogeneous components typically organised in groups.
These entities interact with each other by adapting their behaviour to pursue individual or collective goals. In these systems, the
distribution of these entities determines a space that can be either physical or logical. The former is defined
in terms of a physical relation among components. The latter depends on logical relations, such as being part of the same group. In this context, specification and verification of spatial properties play a fundamental role in supporting the design of systems and predicting their behaviour. %
For this reason, different tools and techniques have been proposed to specify and verify the properties of space, mainly described as graphs. Therefore, the approaches generally use model spatial relations to describe a form of proximity among pairs of entities.
Unfortunately, these graph-based models do not permit considering relations among more than two entities that may arise when one is interested in describing aspects of space by involving \emph{interactions among groups of entities}.
In this work, we propose a spatial logic interpreted on \emph{simplicial complexes}. These are topological objects, able to represent surfaces and volumes efficiently that generalise graphs with higher-order edges. We discuss how the satisfaction of logical formulas can be verified by a correct and complete model checking algorithm, which is linear to the dimension of the simplicial complex and logical formula. The expressiveness of the proposed logic is studied in terms of the spatial variants of classical \emph{bisimulation} and \emph{branching bisimulation} relations defined over simplicial complexes.
  \end{abstract}

\maketitle

\section*{Introduction}\label{S:one}

Collective Adaptive Systems (CAS) often consist of a huge amount of heterogeneous components or entities controlling smart devices~\cite{holzl2008engineering}. These entities are typically arranged in groups and interact with each other to pursue individual or collective goals~\cite{ferscha2015collective}. %
The organization of the entities and the relations among them determine a space that can be either physical or logical. The former depends on the physical position of components in the environment. The latter is related to logical relations, such as being part of the same group or working in the same team. Both these relations may affect the behaviour of each single component as well as the one of the whole system.
To understand the behaviour of CAS, one should rely on formalisms that are able to describe the \emph{spatial structure} of components and on formal tools that support specification and verification of the required spatial properties.
These analyses should be performed from the beginning of the system design.

They facilitate forecasting the impact of the different choices throughout the whole development phase.
Let us consider, for instance, a bike sharing system where resources, namely the bike stations, must be placed in a city.  In this context, it is crucial to allocate resources to avoid having stations too close or too far away.
Moreover, this allocation should consider how \emph{popular} a given zone is by taking into account the position of the nearest points of interest.
The \emph{spatial properties} can be used to evaluate how \emph{popular} is a given area, and can be used to estimate the number of requests for bikes that can be received.
This is because stations in \emph{popular areas} should receive a larger number of requests and for this reason they should be larger than the ones placed in other areas of the city.

 Recently, a considerable amount of work has been proposed focused on the so-called spatial logics~\cite{aiello2007handbook}. These are logical frameworks that give a spatial interpretation to classical temporal and modal logics. Modalities may be interpreted on topological spaces, as proposed by Tarski, or on discrete models (graphs).
For the discrete models, we can refer here to the  notable works by Rosenfeld~\cite{kong1989digital,rosenfeld1979digital}, Galton~\cite{galton2014discrete,galton2003generalized,galton1999mereotopology}, and Smyth and Webster~\cite{smyth2007discrete}.
Recently, Ciancia \textit{et al.}~\cite{ciancia2014specifying} proposed a methodology to verify properties depending upon physical space by defining an appropriate spatial logic whose spatial modalities rely on the notion of \emph{neighbourhood}.
This logic is equipped with spatial operators expressing properties like \emph{surround} and \emph{propagation}.
In~\cite{massink2017model} the approach of~\cite{ciancia2014specifying}  has been extended to handle a set of points in space connected groups of points, rather than points in isolation.
These methodologies have been used to support specification and analysis of a number of scenarios~\cite{MBLSF21, TPGN18, PGPTBFHN17, TKG16}.

However, all the above mentioned approaches do not explicitly consider surfaces or volumes and do not take into account \emph{higher-order relationships} among space entities.
Indeed, they mainly focus on properties among individual entities.
This is due to the fact that graphs are used to describe the underlying spatial models.
Although graphs are a successful paradigm, they cannot explicitly describe groups of interactions.
Indeed, simple dyads (edges) cannot formalise higher-order

relationships.
A possible solution is to obtain information on higher-order interactions in terms of low-order interactions, obtained using clique~\cite{palla2005uncovering} or block detection~\cite{karrer2011stochastic} techniques.
However, the use of higher-order relationships in the system representation requires a more sophisticated mathematical tool: either {\emph{simplicial complexes}~\cite{spanier1989algebraic}  or \emph{hypergraphs}~\cite{berge1973graphs}.

Simplicial complexes are a collection of simplices, i.e., nodes, links, triangles, tetrahedra, \ldots.  Each of them, $\sigma =[v_0, \ldots, v_k]$, is characterised by a dimension $k$, and it can be interpreted as an interaction among $k+1$ entities.  For example, a triangle $[v_0, v_1, v_2]$ is a simplex of dimension $2$, or a $2$-simplex, composed of $3$ vertices and represents a relation among three entities. A characteristic feature of each simplex $\mathcal{K}$ is that all subsets of $\mathcal{K}$ must also be simplices. Thus, the triangle $[v_0, v_1,v_2]$  represents a relation among three entities, $v_0$, $ v_1$, and $v_2$, and implies all the relations between two entities ($[v_0, v_1], [v_0, v_2], [v_1, v_2$]) and the single relations, i.e., the nodes $[ v_0], [v_1], [v_2]$. In this way, simplicial complexes differ from hypergraphs.

A hypergraph consists of nodes set and a set of hyper-edges $H$, which specify the nodes involved in each interaction. Therefore, a hypergraph can include a relation among three entities without any requirement on the existence of pairwise relations. Such property 
comes with additional complexity in treating them. Figure~\ref{higherorderrepresentation} graphically summarises the features and the differences of simplicial complexes and hypergraphs considering the set of relations $\{[v_0, v_1, v_2], [v_1, v_3], [v_1,v_4], [v_3, v_4] \}$.  We observe that the constraint on the sub-complex is intrinsically satisfied by the concept of proximity and spatiality. For example, three neighbouring points in space imply that each pair of them are neighbours.

\begin{figure}
\centering
\includegraphics[scale=0.7]{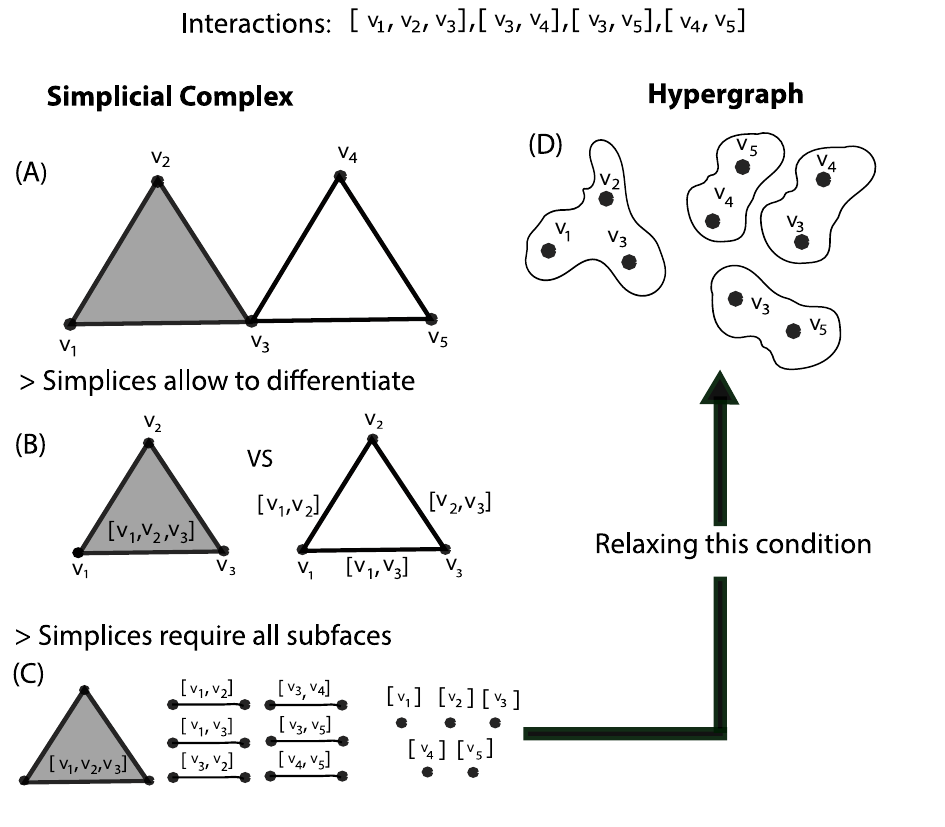}
\caption{Representations of higher-order interactions. The set of interactions $[v_1, v_2, v_3]$, $[v_3,v_4]$, $[v_4,v_5]$, $[v_3,v_5]$ is represented using a simplicial complex (A). Simplicial complexes allow to discriminate between higher order interactions and sums of low-order ones (B). They require the presence of all possible subsimplices (C). Relaxing this condition effectively implies moving from simplices to hypergraphs (D), which are the most general and less constrained representation of higher-order interactions.}%
\label{higherorderrepresentation}
\end{figure}

In this work, we define a spatial logic on simplicial complexes that is able to verify spatial properties on volumes and surfaces in the physical spaces or higher-ordered relations in the case of logical spaces.
 Similar to Ciancia \textit{et al.}, we define two logic operators: \emph{neighborhood}, $\mathcal{N}$, and \emph{reachability}, $\mathcal{R}$, that are reminiscent of the standard \emph{next} and \emph{until} operators of CTL~\cite{CE81,huth2004logic}.

 Interpretation of $\mathcal{N}$ and $\mathcal{R}$ operators is straightforward. A simplex $\sigma$ satisfies $\mathcal{N}\varphi$ if it is in the ``neighborhood'' of another simplex satisfying $\varphi$. This operator recalls the next $X$ operator in CTL:\@ $X(\varphi)$ holds if $\varphi$ is satisfied at the next state. A simplex $\sigma$ satisfies $\varphi_1 \mathcal{R}\varphi_2$ if it satisfies the property $\varphi_2$ or it satisfies $\varphi_1$ and can ``reach'' a simplex that satisfies $\varphi_2$ traversing by a set of simplices satisfying $\varphi_1$.
This operator recalls the until operator $U$ in CTL:\@ $\varphi_1 U \varphi_2$ holds for a path if there is some state along the path for which $\varphi_2$ holds, and $\varphi_1$ holds in all states prior to that state.
To support model checking of spatial properties, we have also defined a procedure that permits checking if a formula is satisfied by a given model. This procedure is proved to be correct and complete, and that it is linear with the dimensions of the considered model and of the logical formula.
 Finally, to study the expressiveness of the proposed logic, we have proposed a variant of (strong) \emph{bisimulation} and \emph{branching bisimulation}. Two fragments of the proposed logic are  identified that fully characterise the two proposed equivalences.

The paper is organized as follows. In Section~\ref{MotivatingExample}, we present two simple examples that motivate our work. Section~\ref{Simplices} recalls some background concepts regarding the simplices and simplicial complexes and the adjacent relations among simplices.  Section~\ref{SpatialLogics} presents the syntax and the semantics of Spatial logic for Simplicial Complexes. In Section~\ref{sec:ModelChecking}, the model checking algorithms for the spatial logic interpreted on
simplicial models are presented. In Section~\ref{Bisimulations}, we study the expressive power of our spatial logic by introducing two equivalences on simplicial models. The two relations are bisimulation and branching bisimulation. In Section~\ref{sec:RelatedWork}, we present an overview of the existing logics dealing with spatial aspects of systems. The paper ends with some conclusions and future work in Section~\ref{Conclusion}.

\section{Motivating Examples}%
\label{MotivatingExample}
In this section, we propose two motivating examples for our spatial logic. In the first one, simplicial complexes are used to represent higher-order interactions among entities. In the second, simplicial complexes are used to represent a physical space.

\subsection{Scientific collaborations}%
\label{MotivatingExample:sc}

The first example we consider is a network of \emph{scientific collaborations}. Given a set of authors $\mathcal{A}=\{a_1,\ldots,a_n\}$ and a set of publications $\mathcal{P}=\{p_1,\ldots,p_k\}$, we want to study the structure of research groups and their collaborations.
Here, we say that a set of authors is a \emph{research group} (or simply a \emph{group}) if they have co-authored at least a paper.

We want to study these collaborations in order to understand which categories of researchers are more collaborative than others or able to interact with other disciplines. For instance, one could be interested in the identification of
\begin{itemize}
\item  Q1.\ the \emph{groups} containing authors of at least a paper on topic ``A'',
\item  Q2.\ chains of \emph{groups} on topic ``A'' that leads to work on topic ``B''.
\end{itemize}

\subsection{Emergency Rescue}%
\label{MotivatingExample:sc:ee}

Let us consider a scenario where in a given area an accident occurred that caused the emission of dangerous gasses or radiations. To identify the dangerous zones in the area a number of sensors are spread via a helicopter or an airplane.
Each sensor $s$ is able to measure the degree of hazard in a radius $\delta$ and it can also perceive if a victim, identified by a cross in Figure~\ref{imm:figbuilding},  is in its surrounding. In what follows we let $A_{s}$ denote the area observed by sensor $s$ (see Figure~\ref{imm:figbuilding}).
Sensors can be used by a rescue team to identify the \emph{safe paths} and \emph{zones} in the area to allow them to identify safer routes to reach victims.
In other words, we are interested in reaching ``victims'' only through ``safe'' areas, i.e., through areas characterised by a percentage of toxicity less than a certain threshold.

\newcommand{\picsensor}[5]{
	\node (#1) [sensor,label=#2] at (#3,#4) {}; 
	\draw[black,dotted] (#3,#4) circle (#5);
}
\newcommand{\picradiation}[3]{
	\draw[red!10,fill=red!10] (#1,#2) circle (#3);
}

\tikzset{cross/.style={cross out, draw,
         minimum size=2*(#1-\pgflinewidth), 
         inner sep=0pt, outer sep=0pt}}

\begin{figure}
\centering
\begin{tikzpicture}[scale=1.12,sensor/.style={shape=circle,draw,fill=black,minimum size=1mm,inner sep=0.1}]

	\draw (-.5,-.5) -- (5.55,-.5) -- (5.55,3.53) -- (-.5,3.53) -- (-.5,-.5);
	\picradiation{1.25}{2.85}{.5}
	\picradiation{1.75}{2.75}{.5}
	\picradiation{2.25}{2.85}{.5}
	\picradiation{2.75}{2.65}{.5}
	\draw (4.25,1.25) node[cross=3pt,rotate=10,blue] {};
	\picsensor{s1}{$s_1$}{.75}{.75}{1}
	\picsensor{s2}{$s_2$}{1.25}{1.25}{1}
	\picsensor{s3}{$s_3$}{2.15}{1.25}{1}
	\picsensor{s4}{$s_4$}{1.60}{1.75}{1}
	\picsensor{s5}{$s_5$}{2.60}{.9}{1}
	\picsensor{s6}{$s_6$}{2.30}{2.0}{1}
	\picsensor{s7}{$s_7$}{1.95}{2.5}{1}
	\picsensor{s8}{$s_8$}{3.2}{2.0}{1}
	\picsensor{s9}{$s_9$}{3.2}{.5}{1}
	\picsensor{s10}{$s_{10}$}{3.8}{.9}{1}
	\picsensor{s11}{$s_{11}$}{4}{1.75}{1}
	\picsensor{s12}{$s_{12}$}{4.5}{1.5}{1}
\end{tikzpicture}
\caption{A representation of some sensors in area to identify dangerous zones, where each $s_{i}$ represents a sensor while the blue cross identifies the victim.}%
\label{imm:figbuilding}
\end{figure}
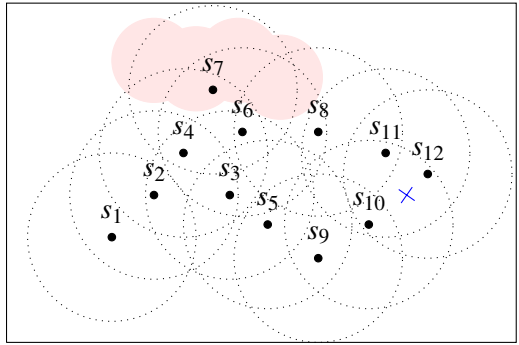

\section{Simplices and  Simplicial Complexes}%
\label{Simplices}
Simplices~\cite{munkres2018elements} are a generalisation of the notion of a triangle or tetrahedron to arbitrary dimensions. The dimension of simplex $\sigma$, $\dim(\sigma)$, is defined as the number of vertices of $\sigma$ minus one, $\dim(\sigma) = |\sigma|-1$.
We let $[v_1,\ldots, v_{k+1}]$ denote the $k$-simplex $\sigma$ composed of the vertices $\{v_1, \ldots, v_{k+1}\}$.
Often, the term $k$-simplex is used to refer to a simplex of dimension $k$.
For example,  a $0$-simplex is a point; a $1$-simplex is a line, connecting two points; a $2$-simplex is a triangle, composed by three lines; while a $3$-simplex is a solid tetrahedron, delimited by four triangles.
Their representation is shown in Figure~\ref{simplices}.

Each $k$-simplex is formed by $k+1$ simplices of dimension $(k-1)$.
Given a $k$-simplex $\sigma=[v_1,\ldots, v_{k+1}]$, we say that a $k'$-simplex $\sigma'=[w_1,\ldots, w_{k'+1}]$ ($k'<k$) is a \emph{face} of $\sigma$ if and only if $\{w_1,\ldots, w_{k'+1}\} \subset \{v_1, \ldots, v_{k+1}\}$.
Moreover, given two simplices $\sigma_1$ and $\sigma_2$ we let $\sigma_1\cap \sigma_2$ denote the simplex composed of the vertices occurring in both $\sigma_1$ and $\sigma_2$, while we will write $\sigma_1 \subseteq \sigma_2$ whenever the vertices in $\sigma_1$ are also vertices in $\sigma_2$.

\begin{figure}[h!]
\centering
\hspace{2.5mm}
\begin{tikzpicture}[scale=.12]
\filldraw (1,1) circle[radius=12pt];
\end{tikzpicture}
\hspace{5.5mm}
\begin{tikzpicture}[scale=.3]
\begin{scope}[every node/.style={fill=black,ultra thick,circle,scale=0.3}]
\node (n3) at (7,1) {1};
\node (n4) at (1,1) {2};
\end{scope}
   \begin{scope}[every edge/.style={draw=black, thick,scale=0.2}]
  \draw  (n3) edge node{} (n4);
 \end{scope}
\end{tikzpicture}
\hspace{5.5mm}
\begin{tikzpicture}[scale=.3]
\begin{scope}[every node/.style={fill=black,circle,scale=0.3}]
\node (n3) at (7,1) {1};
  \node (n4) at (1,1) {2};
  \node (n5) at (3.5,3) {2};
\end{scope}

   \begin{scope}[every edge/.style={draw=black,thick,scale=0.2}]

  \draw  (n3) edge node{} (n4);
    \draw  (n4) edge node{} (n5);
  \draw  (n5) edge node{} (n3);

 \end{scope}
 \draw[pattern=north west lines, pattern color=black]  (n3.center) -- (n4.center) -- (n5.center) -- cycle;
\end{tikzpicture}
\hspace{5.5mm}
\begin{tikzpicture}[scale=.3]

\begin{scope}[every node/.style={fill=black, thick, circle,scale=0.3}]
\node (n3) at (7,1) {1};
  \node (n4) at (1,1) {2};
  \node (n5) at (4,3) {2};
  \node (n6) at (4,-3) {2};

\end{scope}
   \begin{scope}[every edge/.style={draw=black, thick,scale=0.1}]

  \draw  (n3) edge node{} (n4);
    \draw  (n4) edge node{} (n5);
  \draw  (n5) edge node{} (n3);
  \draw  (n6) edge node{} (n4);
  \draw  (n6) edge node{} (n5);
  \draw  (n6) edge node{} (n3);
\end{scope}
\draw[pattern=north west lines, pattern color=black]  (n3.center) -- (n4.center) -- (n5.center) -- cycle;
\draw[pattern=north west lines, pattern color=black]  (n6.center) -- (n5.center) -- (n3.center) -- cycle;
\draw[pattern=north west lines, pattern color=black]  (n3.center) -- (n4.center) -- (n6.center) -- cycle;

\end{tikzpicture}
\caption{From left to right simplices of dimension zero, one, two and three are shown.}%
\label{simplices}
\end{figure}
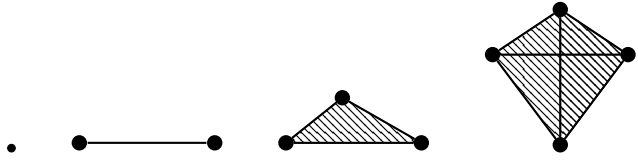

In this work, we are interested in sets of simplices that are closed under taking faces and that have no improper intersections, the so-called \textit{simplicial complex}. Formally,

\noindent

\begin{defi}[Simplicial complex]
A simplicial complex $\mathcal{K}$ is a collection of simplices, such that
\begin{enumerate}
\item every face of a simplex of $\mathcal{K}$ is also in $\mathcal{K}$;
\item the intersection of any two simplices $\sigma_i$, $\sigma_j$ of $\mathcal{K}$ is either $\emptyset$ or a face of both  $\sigma_i$ and $\sigma_j$.
\end{enumerate}%
\label{def:simplicialcomplex}
\end{defi}

\noindent
The dimension of  a simplicial complex $\mathcal{K}$ is the maximum of the dimensions of all simplices in $\mathcal{K}$.
We observe that a  simplicial complex of dimension $1$ is a graph.

The collection in Figure~\ref{SimplicialComplex}-(A) composed of $7$ vertices ($0$-simplices), $10$ edges ($1$-simplices), $5$ triangles ($2$-simplices) and a tetrahedron ($3$-simplex) is a simplicial complex, while the collection illustrated in Figure~\ref{SimplicialComplex}-(B) violates the definition of a simplicial complex because the intersection of the two triangles does not consist of a complete edge. 

\begin{figure}[htp]
\centering
\begin{tabular}{c@{\hspace{2cm}}c}
\begin{tikzpicture}[scale=0.67]
    \tikzstyle{point}=[circle,thick,draw=black,fill=black,inner sep=0pt,minimum width=4pt,minimum height=4pt]
    \node (a)[point] at (0,0) {};
    \node (b)[point] at (2.5,0) {};
    \node (c)[point] at (2,2) {};

    \begin{scope}[yshift=2cm]
    \node (d)[point] at (1,1) {};
    \node (e)[point] at (0,2) {};
    \node (f)[point] at (2.4,2) {};
    \end{scope}
    \node (p)[point,label={[label distance=0cm]5:$ $}] at (1.5,0.5) {};

    \draw[pattern=north east lines] (a.center) -- (p.center) -- (b.center) -- cycle;
    \draw[pattern=north west lines] (a.center) -- (p.center) -- (c.center) -- cycle;
    \draw[pattern=vertical lines]   (b.center) -- (p.center) -- (c.center) -- cycle;
    \draw[pattern=dots] (d.center) -- (e.center) -- (f.center) -- cycle;
    \draw (p.center) -- (d.center);
\end{tikzpicture}
&
\begin{tikzpicture}
    \tikzstyle{point}=[circle,thick,draw=black,fill=black,inner sep=0pt,minimum width=4pt,minimum height=4pt]
   \node (a)[point] at (0,0) {};
    \node (b)[point] at (3,0) {};
    \node (c)[point] at (2,2) {};

    \begin{scope}[yshift=2cm]
    \node (d)[point] at (1,1) {};
    \node (e)[point] at (0,2) {};
    \node (f)[point] at (3.5,2) {};
    \end{scope}

    \begin{scope}[yshift=2cm]
    \node (g)[point] at (2.5,2) {};
    \node (h)[point] at (4.5,2) {};
    \node (i)[point] at (3,1) {};
    \end{scope}

    \node (p)[point,label={[label distance=0cm]5:$ $}] at (1.5,0.5) {};

    \draw[pattern=north east lines] (a.center) -- (p.center) -- (b.center) -- cycle;
    \draw[pattern=north west lines] (a.center) -- (p.center) -- (c.center) -- cycle;
    \draw[pattern=vertical lines]   (b.center) -- (p.center) -- (c.center) -- cycle;
    \draw[pattern=dots] (d.center) -- (e.center) -- (f.center) -- cycle;
    \draw (p.center) -- (d.center);
     \draw[pattern=north east lines] (g.center) -- (h.center) -- (i.center) -- cycle;
\end{tikzpicture}
\\
(A) & (B)
\end{tabular}
\caption{In (A), a simplicial complex; in (B), a collection of simplices that is not a simplicial~complex.}%
\label{SimplicialComplex}
\end{figure}
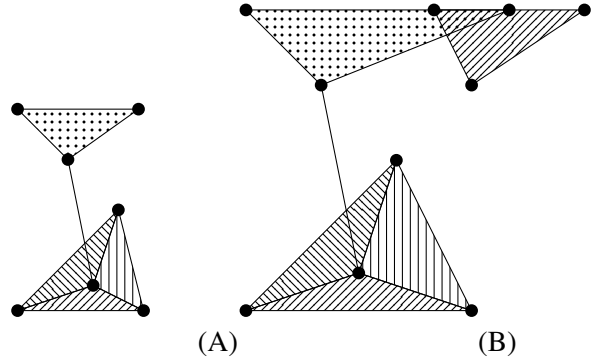
The following two examples show how both the scenarios considered in Section~\ref{MotivatingExample} can be described in terms of simplicial complexes.


\begin{exa}%
\label{ex:simplicial_definitions:sc}
To model the higher-order relation ``co-authorship group of $k$ researchers'' considered in Section~\ref{MotivatingExample:sc} we can use the approach of q-analysis to study social systems~\cite{atkin1974mathematical}. Therefore, we make use of a geometric interpretation of the relationships between entities and events.  The nodes ($0$-simplices) identify the authors, links ($1$-simplices) represent a pair of co-authors, while each $k$-simplex with $k$ more than one represents the relations  ``co-authorship group of $k$ researchers''.  Thus, the simplicial complex in Figure~\ref{higherorderrepresentation} (A) represents four groups of co-authors, one of them determined by $v_1$, $v_2$, and $v_3$, and three others consisting of two authors.
Let $\mathcal{K}_{\mathcal{A}}$ be the simplicial complex that formalises a group of authors $\mathcal{A}=\{a_1,\ldots,a_n\}$ of papers $\mathcal{P}=\{p_1,\ldots,p_w\}$. Each simplex of the form  $[a^i_1,\ldots,a^i_k]$ is an element of $ \mathcal{K}_{\mathcal{A}}$ if and only if there exists a paper $p_j\in \mathcal{P}$ such that $\{a_1,\ldots,a_k\}$ are among the authors of $p_j$.
For instance, in Figure~\ref{imm:fig3} a network of six authors is depicted that are arranged in four groups: $[a_1, a_2, a_3]$, $[a_2, a_4, a_5]$, $[a_4, a_6]$ and $[a_5, a_6]$.
\end{exa}


\begin{figure}[htp]
\centering
\includegraphics[scale=0.25]{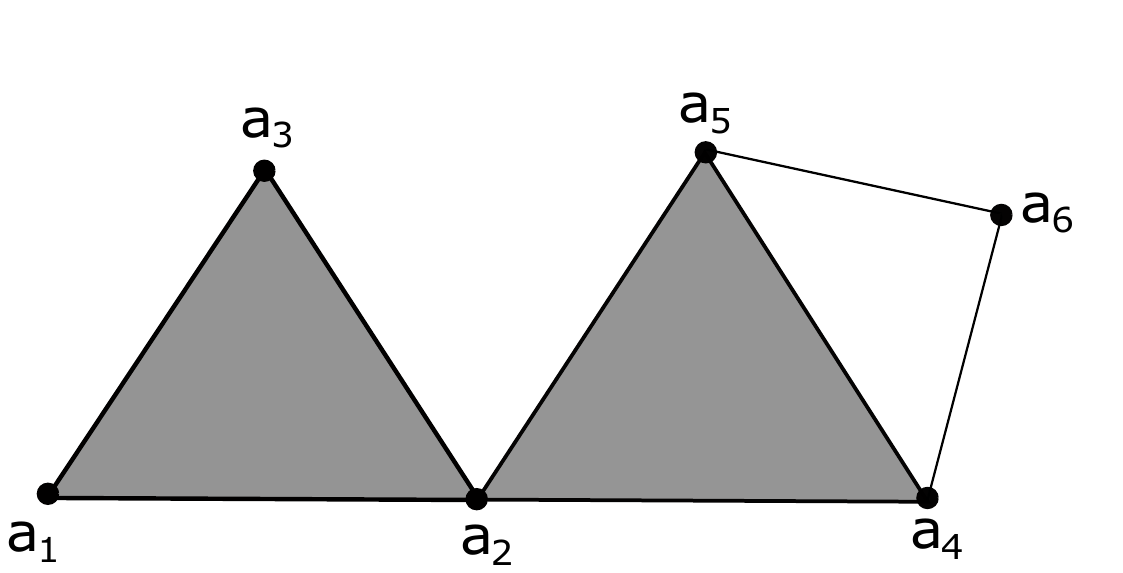}
\caption{An network of scientific collaborations.}%
\label{imm:fig3}
\end{figure}

\begin{exa}%
	\label{ex:simplicial_definitions:ee}

The sensors located in the area considered in Section~\ref{MotivatingExample:sc:ee} can be represented via a simplicial complex where:
\begin{itemize}
	\item 0-simplices (points) corresponds to sensors;
	\item 1-simplices (edges) consists of the set of $[s_i,s_j]$ such that $\{s_i,s_j\}\subseteq A_{s_i}\cap A_{s_j}$;
	\item 2-simplices (areas) consists of the set of $[s_i,s_j,s_k]$ such that  $\{s_i,s_j,s_k\}\subseteq A_{s_i}\cap A_{s_j} \cap A_{s_k}$.
\end{itemize}
We let $\mathcal{K}_{\mathcal{E}}$ denote the simplicial complex described above. A representation is depicted in Figure~\ref{imm:complex}.

\begin{figure}[htp]
\centering
\begin{tikzpicture}[scale=1.12,sensor/.style={shape=circle,draw,fill=black,minimum size=1mm,inner sep=0.1}]

	\draw (-.5,-.5) -- (5.55,-.5) -- (5.55,3.53) -- (-.5,3.53) -- (-.5,-.5);
	\picradiation{1.25}{2.85}{.5}
	\picradiation{1.75}{2.75}{.5}
	\picradiation{2.25}{2.85}{.5}
	\picradiation{2.75}{2.65}{.5}
	\draw (s1) -- (s2);
	\draw [fill=black!10] (s2.center) -- (s3.center) -- (s4.center) -- (s2.center);
	\draw (s3) -- (s5);
	\draw [fill=black!10] (s3.center) -- (s4.center) -- (s6.center) -- (s3.center);
	\draw [fill=black!10] (s4.center) -- (s6.center) -- (s7.center) -- (s4.center);
	\draw (s6) -- (s8);
	\draw (s5) -- (s9);
	\draw (s9) -- (s10);
	\draw (s8) -- (s11);
	\draw [fill=black!10] (s10.center) -- (s11.center) -- (s12.center) -- (s10.center);

	\draw (4.25,1.25) node[cross=3pt,rotate=10,blue] {};

	\picsensor{s1}{$s_1$}{.75}{.75}{1}
	\picsensor{s2}{$s_2$}{1.25}{1.25}{1}
	\picsensor{s3}{$s_3$}{2.15}{1.25}{1}
	\picsensor{s4}{$s_4$}{1.60}{1.75}{1}
	\picsensor{s5}{$s_5$}{2.60}{.9}{1}
	\picsensor{s6}{$s_6$}{2.30}{2.0}{1}
	\picsensor{s7}{$s_7$}{1.95}{2.5}{1}
	\picsensor{s8}{$s_8$}{3.2}{2.0}{1}
	\picsensor{s9}{$s_9$}{3.2}{.5}{1}
	\picsensor{s10}{$s_{10}$}{3.8}{.9}{1}
	\picsensor{s11}{$s_{11}$}{4}{1.75}{1}
	\picsensor{s12}{$s_{12}$}{4.5}{1.5}{1}
\end{tikzpicture}
\caption{ The simplicial complex representation of the area with some sensors shown in Figure~\ref{imm:figbuilding}.}%
\label{imm:complex}
\end{figure}
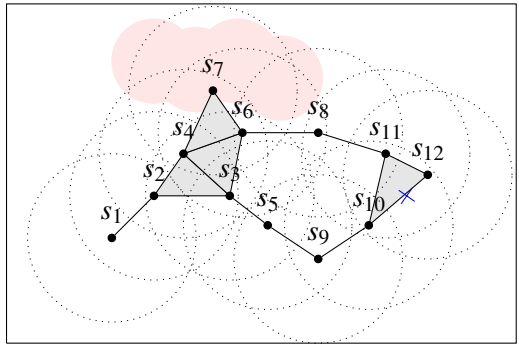

\end{exa}

In graph theory, two nodes are adjacent if they are linked by an edge. Such concept can be extended in simplicial complexes. In the literature, two relations are used to characterise the adjacency of simplicial complexes: \textit{lower} and \textit{upper} adjacency~\cite{tahbaz2010distributed,maletic2012combinatorial}.

\begin{defi}[Lower adjacency] Let  $\mathcal{K}$ be a simplicial complex and let $\sigma_i, \sigma_j $ be two  distinct $k$-simplices in $\mathcal{K}$. Then the two $k$-simplices are \textit{lower adjacent} if they share a common face. That is, $\sigma_i$ and $\sigma_j$ are \textit{lower adjacent} if and only if there is a $(k -1)$-simplex $\sigma$ such that $\sigma \subset \sigma_i$ and $ \sigma \subset \sigma_j$. We denote lower adjacency by $\sigma_i \smile \sigma_j$.%
\label{def:lower}
 \end{defi}

For instance, in the simplicial complex $\mathcal{K}$ of Figure~\ref{imm:fig3}, the $1$-simplices $[a_1, a_3] $ and $[ a_3, a_2]$ are lower adjacent because the $0$-simplex $[a_3]$ is their common face and we can write $ [a_1, a_3] \smile [a_3, a_2]$. However, $[a_1,a_2,a_3]$ and $[a_2,a_4,a_5]$ are not lower adjacent although they have the common $0$-simplex $[a_2]$. In fact, to be lower adjacent they need to share a common $1$-simplex.

\begin{defi}[Upper adjacency]%
 \label{def:upper}
 Let  $\mathcal{K}$ be a simplicial complex and let $\sigma_i, \sigma_j $ be two distinct $k$-simplices in $\mathcal{K}$. Then, the two $k$-simplices are upper adjacent if they are both faces of the same common $(k + 1)$-simplex. That is, $\sigma_i$ and $\sigma_j$ are upper adjacent if and only if there is a $(k+1)$-simplex $\sigma$ such that $\sigma_i \subset \sigma$ and $\sigma_j \subset \sigma$. We denote the upper adjacency by $\sigma_i \frown \sigma_j$.
\end{defi}

 In the simplicial complex in Figure~\ref{imm:fig3}, the $1$-simplices $[a_1, a_3]$ and $[a_1, a_2]$ are upper adjacent because they are both faces of the $2$-simplex $[a_1, a_2, a_3]$. So we can write $[a_1, a_3] \frown [a_1,a_2]$}. However, $[a_1,a_2,a_3]$ is not upper adjacent to any other simplex as it is not part of any $3$-simplices. Also note that $[a_5]$ and $[a_6]$ are upper adjacent because they are both faces of $[a_5, a_6]$. Hence, two $0$-simplices are upper adjacent if they are both faces of a $1$-simplex which is identical to saying that two nodes are adjacent if they are connected by an edge in the graph. Note that upper adjacency of $0$-simplices corresponds to the classical graph adjacency.
 Since simplicial complexes are composed of sets of simplices closed under taking faces, we propose to generalise the concept of adjacency in terms of common faces (simplices) as follows.

  \begin{defi}[Spatial adjacency] Let  $\mathcal{K}$ be a simplicial complex, and let $\sigma_i$  and $\sigma_j$ be two distinct simplices of dimension $k_i$ and $k_j$, respectively, in $\mathcal{K}$. Then, the two simplices are spatial adjacent  if they share at least a face, that is $\sigma_i \cap \sigma_j  \neq \emptyset$. We denote the spatial adjacency by $\sigma_i  \multimapdotboth \sigma_j$.
\end{defi}

We can observe that a simplex $\sigma_i$ is \emph{spatial adjacent} to any of its faces $\sigma_j$.
In the simplicial complex in Figure~\ref{imm:fig3}, $[a_1,a_2,a_3]$ and $[a_2,a_4,a_5]$ are spatial adjacent because the $0$-simplex $[a_2]$ which is a face of both of them. Note that these two simplices are not lower adjacent. In fact, differently from \emph{lower} and \emph{upper} adjacency, in the definition of spatial adjacency we have not any requirement on the dimensions of involved simplices.
Note that \emph{lower} and \emph{upper adjacency} are useful whenever one is interested in considering simplices of a given dimension. For instance, in Example~\ref{ex:simplicial_definitions:ee} these relations permit identifying connected areas or vertices. On the contrary, \emph{spatial adjacency} allows us to describe connections between simplices of different size. This is the case of simplicial complexes of Example~\ref{ex:simplicial_definitions:sc} where one can consider connections among \emph{groups} of different size.

\begin{prop}

Any two upper adjacent $k$-simplices, with $k>0$, of a simplicial complex $\mathcal{K}$  are also  lower adjacent.%
   \label{up-low}
  \end{prop}

 \begin{proof}  Let $\sigma_i$ and $\sigma_j$ be two upper adjacent $k-$simplices in $\mathcal{K}$, with $k>0$.
From Definition~\ref{def:upper}, there exists a $(k+1)$-simplex $\tau = [v_1, v_2, \dots, v_{k +2}]$  such that $\sigma_i \subset \tau $ and $\sigma_j \subset \tau$.
Without loss of generality, we can assume that $\sigma_i =[ v_1 , v_2, \dots, v_{k+1}]$  and $\sigma_i = [v_2 , \dots, v_{k +1}, v_{k +2}] $.
Let $\sigma'= \sigma_i \cap \sigma_j$.
It is easy to see that $\sigma'$ is the $(k-1)$-simplex $[ v_2 ,\dots, v_{k +1} ]$ in $\mathcal{K}$ (see Definition~\ref{def:simplicialcomplex}).
Hence, $\sigma_i$ and $\sigma_j$ are lower adjacent since $\sigma'\subset \sigma_i$ and $\sigma'\subset \sigma_j$.
\qedhere
 \end{proof}

  \begin{prop}%
  \label{thlower}
  Any two lower adjacent $k$-simplices of a simplicial complex $\mathcal{K}$  are spatially adjacent.
\end{prop}

\begin{proof}  Suppose $\sigma_i$ and $\sigma_j$ are lower adjacent simplices in $\mathcal{K}$.
By Definition~\ref{def:lower}, the dimension of $\sigma_i$ is equal to the dimension of $\sigma_j$, that we impose equal to $k$, and there exists $\sigma \in \mathcal{K}$ such that $\dim(\sigma) = k-1$, $\sigma \subset \sigma_i$ and $\sigma \subset \sigma_j$.
The claim follow directly from the fact that $\sigma\subseteq \sigma_i\cap \sigma_j$.
\qedhere
\end{proof}

\begin{prop}%
\label{thupper}
Any two upper adjacent $k$-simplices with $k>0$ of a simplicial complex $\mathcal{K}$  are spatially adjacent.
 \end{prop}

\begin{proof} Suppose $\sigma_i$ and $\sigma_j$ are upper adjacency in $\mathcal{K}$. From the definition of upper adjacency, it follows that the dimension of $\sigma_i$ is equal to the dimension of $\sigma_j$, that we set equal to $k$, and exist $\sigma \in \mathcal{K}$ such that $\dim(\sigma) = k+1$, $\sigma_i \subset \sigma$ and $\sigma_j \subset \sigma$. Since $\sigma_j$ and $\sigma_i$ are two $k$-faces of the same $(k+1)$-simplex, they are lower adjacent (Proposition~\ref{up-low}). Therefore, from Proposition~\ref{thlower}, $\sigma_i$ and $\sigma_j$ are spatial adjacent.
\qedhere
\end{proof}

\section{Spatial logics for Simplicial Complexes}%
\label{SpatialLogics}

In this section, we introduce Spatial Logic for Simplicial Complexes (SLSC). The logic features boolean operators,  a ``one step''  modality referred to as  \textit{Neighbourhood} and denoted by $\mathcal{N}$, and a binary spatial operator, called  \textit{Reachability} and denoted by  $\mathcal{R}$, that are evaluated upon a set of simplices.
Assume a finite or countable set $P$ of atomic propositions.

\begin{defi}[Simplicial Model]
A simplicial model is a triple $\mathcal{M} = ( \mathcal{K}, P, \nu)$, where   $\mathcal{K}$  is a simplicial complex, $P$ is a set of atomic propositions and  $\nu: P \to \mathcal{K}$ is a valuation function  that assigns to each atomic proposition the set of simplices where the proposition holds.
For any $\sigma_1, \sigma_2\in \mathcal{K}$, we will write $\sigma_{1}\equiv_{\nu} \sigma_2$ if and only if for any $a\in P$, $\sigma_1\in \nu(a)\Leftrightarrow \sigma_2\in \nu(a)$. Moreover, for any $\sigma\in \mathcal{K}$, we let $\lambda(\sigma)$ denote $\{a\in P| \sigma\in \nu(a) \}$.
 \end{defi}

\begin{exa}%
\label{ex:scm:sc}
To reason about simplicial complex $\mathcal{K}_{\mathcal{A}}$ of Example~\ref{ex:simplicial_definitions:sc}, we let $\mathcal{M}_{\mathcal{A}}=(\mathcal{K}_{\mathcal{A}},P_{\mathcal{A}}, \nu_{\mathcal{A}})$, where $P_{\mathcal{A}}=\{A_1,\ldots,A_w\}$ consists of the set of topics of considered papers, while $\nu_{\mathcal{A}}$ associates with a simplex $[a^i_1,\ldots,a^i_k]$ a topic $A_j$ if and only if $\{a^i_1,\ldots,a^i_k\}$ co-authored a paper with topic $A_j$.
\end{exa}

\begin{exa}%
\label{ex:scm:ee}
The definition of the simplicial model representing the motivating example of Section~\ref{MotivatingExample:sc:ee} is based on the set of atomic propositions $P_{\mathcal{E}}=\{\mathsf{safe},\mathsf{unsafe},\mathsf{victim}\}$. Let $\sigma=[s_1,\ldots,s_k]\in\mathcal{K}_{\mathcal{E}}$ (see Example~\ref{ex:simplicial_definitions:ee}). Valuation function $\nu_{\mathcal{E}}$ associates $\sigma$ with the atomic proposition $\mathsf{unsafe}$ if and only if at least one $s_i \in \sigma$ is measuring an \emph{unsafe} a level of toxicity. Conversely, $\sigma$ satisfies atomic proposition $\mathsf{safe}$ whenever no sensors in $\sigma$ is perceiving any hazard. Similarly, the same simplex satisfies $\mathsf{victim}$ if each sensor $s_i$ perceives a victim.
We let the model be $\mathcal{M}_{\mathcal{E}} = (\mathcal{K}_{\mathcal{E}}, P_{\mathcal{E}}, \nu_{\mathcal{E}})$.
\end{exa}
We can now define the logic.
 \begin{defi}[Syntax] The syntax of SLSC is defined by the following grammar, where $ a$ ranges over $P$:
\[\phi ::=  a \mid \top \mid \lnot \varphi \mid \varphi_1 \wedge \varphi_2 \mid \mathcal{N} \varphi_1 \mid \varphi_1 \mathcal{R}  \varphi_2  \ .\]
 \end{defi}
Here,  $\top$ denotes \textit{true}, $\lnot$ is negation, $\wedge$ is conjunction, $\mathcal{N}$  is the neighborhood operator and  $\mathcal{R}$ is the reachability operator. We shall now define the interpretation of formulas
\begin{defi}[SLSC semantics]\label{Semantics} Let $\mathscr{C}$ be an element of $\{ \smile, \frown, \multimapdotboth \}$. The set of simplices of a simplicial complex $\mathcal{K}$  satisfying formula $\varphi$, that is indicated with $\llbracket \ \varphi \ \rrbracket_{\mathscr{C}}$, in simplicial model $\mathcal{M} = (\mathcal{K}, P, \nu)$ is defined by the following equations
 \begin{align}
 &  \llbracket \ a  \ \rrbracket_{\mathscr{C}}= \nu(a )\label{ap}  \\
 & \llbracket \  \top \ \rrbracket_{\mathscr{C}}  = \mathcal{K}  \\
 & \llbracket \ \varphi_1 \wedge \varphi_2 \ \rrbracket_{\mathscr{C}}  = \llbracket \ \varphi_1 \ \rrbracket_{\mathscr{C}}  \cap  \llbracket \ \varphi_2 \ \rrbracket_{\mathscr{C}}\label{and} \\
 & \llbracket \ \lnot \varphi \ \rrbracket_{\mathscr{C}}  = \mathcal{K}  \setminus \llbracket \ \varphi \ \rrbracket_{\mathscr{C}}\label{not}\\
 & \llbracket  \ \mathcal{N} \varphi \ \rrbracket_{\mathscr{C}} = \{\sigma_1 \in \mathcal{K}: \exists \ \sigma_2 \in \llbracket \  \varphi \ \rrbracket_{\mathscr{C}} \ \mbox{and} \ \sigma_1 \  \mathscr{C} \  \sigma_2  \}\label{N}\\
  & \llbracket  \  \varphi_1 \mathcal{R} \varphi_2 \ \rrbracket_{\mathscr{C}} = \bigcup_{i=0}^{\infty} R^{i}_{\mathscr{C}} (\llbracket  \  \varphi_1 \ \rrbracket_{\mathscr{C}}  , \llbracket\ \varphi_2 \ \rrbracket_{\mathscr{C}})\\
\mbox{where}\\
 & R^{0}_{\mathscr{C}} (\Sigma_1  , \Sigma_2) = \Sigma_2\\
 & R^{i+1}_{\mathscr{C}} (\Sigma_1  , \Sigma_2) =  \{   \sigma_1  \in \Sigma_1 | \exists \sigma_2  \in R^{i}_{\mathscr{C}} (\Sigma_1  , \Sigma_2).   \sigma_1 \  \mathscr{C} \ \sigma_2  \} \end{align}  \end{defi} 
 Atomic propositions and boolean connectives have the expected meaning. For formulas of the form  $\mathcal{N}\varphi$, the basic idea is that a simplex $\sigma$ satisfies $\mathcal{N}\varphi$ if it is adjacent to another one, $\sigma '$, satisfying the formula $\varphi$. Note that  it is not required that $\sigma$ satisfies $\varphi$.
We can observe that when one considers $0$-simplices and \emph{upper adjacency}, $\mathcal{N}\varphi$ coincides with the standard
\emph{next operator} in CTL~\cite{baier2008principles,blackburn2001}.
 A simplex satisfies $\varphi_1 \mathcal{R} \varphi_2$ if it satisfies $ \varphi_2$ or  it satisfies $ \varphi_1$ and can reach a simplex that satisfies $\varphi_2$ passing through a set of adjacent simplices satisfying $\varphi_1$.  The interpretation of these two operators depends on the considered adjacency relation.
For instance, let us  consider the simplices in Figure~\ref{imm:fig3} and let us assume that simplex $\sigma_1=[a_2,a_4,a_5]$ satisfies a formula $\varphi_1$. We can observe that if one considers \emph{spatial adjacency}, formula $\mathcal{N} \varphi_1$ is satisfied by $\sigma_2=[a_1,a_2,a_3]$ (we have already observed that $\sigma_1 \multimapdotboth \sigma_2$). However, the same formula is not satisfied by $\sigma_2$ if one considers \emph{lower} or \emph{upper adjacency}.
 The specific adjacency relation to use depends on the application context. If one is interested in specifying or verifying properties of simplices of a given dimension, either \emph{upper} or \emph{lower adjacency relation} should be used. On the contrary, if one is interested to study spatial properties based on a weaker form of \emph{spatial connection}, \emph{spatial adjacency} should be used. In the following examples we will show how the considered \emph{adjacency relations} permit reasoning about different aspects of our running examples.

\begin{exa}
Let us consider again the simplicial complex describing the network of scientific collaboration shown in Figure~\ref{imm:fig3}. We can assume that the topics of papers published by the considered authors are $\{ t_1, t_2 \}$.
These are the \emph{atomic propositions} that are associated to our model.
We assume that the collaborations described by the simplices $[a_1, a_2, a_3]$, $[a_2, a_4, a_5]$ regards topic $t_1$ while the topic $t_2$ is associated with the collaborations described by $[a_5, a_6]$, $[a_4, a_6]$.
We can observe that some of the authors, namely $[a_5]$ and $[a_4]$, have written papers on both the topic $t_1$ and $t_2$. This means that, these simplices satisfy both the corresponding atomic propositions.

The SLSC formulas of the form $\mathcal{N}~t_i $ allow us to answer questions Q1 of  Section~\ref{MotivatingExample}, i.e.,  if some co-authors of a given paper have also co-authored a paper on topic $t_i$.
For instance, in our example, the simplex $[a_1,a_2,a_3]\in \llbracket\mathcal{N}~t_1 \rrbracket_{\multimapdotboth}$ because the simplex $ [a_2,a_4,a_5]$ satisfies $t_1$, and it is spatial adjacent to $[a_1,a_2,a_3]$.

Moreover, to answer to question Q2 we can used a SLSC formula of the form $t_1 ~\mathcal{R}~t_2$. Both, the simplices $[a_1,a_2,a_3]$ and $[a_2,a_4,a_5]$ satisfies the formula $t_1 \mathcal{R} t_2$ with respect to $\multimapdotboth$.
Indeed, $[a_2,a_4,a_5]$ satisfies $t_1$ and it is \emph{spatial adjacent} to $[a_4,a_6]$ that satisfies $t_2$. Moreover,  $[a_1,a_2,a_3]$ satisfies $t_1$ and it is spatial adjacent to $[a_2,a_4,a_5]$ that satisfies $t_1 ~\mathcal{R}~t_2$.

\end{exa}

\begin{exa}
	We can use SLSC formulas to select the paths and surfaces in  $\mathcal{K}_{\mathcal{E}}$ that are safe and that can be traversed to reach a victim.
	We have seen in Example~\ref{ex:scm:ee} that atomic proposition $\mathsf{safe}$ is satisfied by the points, segments and surfaces without hazards.
	We say that a simplex is \emph{safer} whenever it is $\mathsf{safe}$ and it is not adjacent with an $\mathsf{unsafe}$ simplex. This property can be specified with the formula $\phi_{safer}=\mathsf{safe}\wedge \neg \left( \mathcal{N} \mathsf{unsafe}\right)$.
	To select the areas that the rescue team can use to reach a victim, the following formula can be used $\phi_{safer}~\mathcal{R}~\mathsf{victim}$.

Note that we can use \emph{lower adjacency} to identify a set of \emph{adjacent surfaces} that can be safely traversed to reach a victim. The border of these zones, which are represented as a set of connected lines, can be selected by using \emph{upper adjacency}. Finally, the use of \emph{spatial adjacency} will give us a complete picture of the safe areas, and their connection, in the arena. According to the definition of our logic, we underline different notions of adjacency cannot be use in the same formula. Therefore, the set of simplexes that satisfies the formula depend on the kind of selected adjacency.
\end{exa}

We observe that SLSC generalises the spatial logic proposed in~\cite{massink2017model}. This logic  provides two spatial operators: a \emph{closure}\footnote{In~\cite{massink2017model} the closure operator of the SLCS is denoted by  $\mathcal{N}$ as well as the next operator of our logic. To avoid confusion, here we indicate the closure operator of the SLCS by $\mathcal{C}$.
}  operator $\mathcal{C}$ and a \emph{surround} operator $\mathcal{S}$. Both these  operators are interpreted over \emph{closure spaces} in terms of application of the closure operators.
When graph-based structures are considered, this closure operator is based on a \emph{binary relation} and the closure consists in a \emph{one-step closure} of the set.
Both these operators can be expressed in our formalism as macro of the $\mathcal{N}$ and $\mathcal{R}$ operators. Indeed, we have that $\mathcal{C}\varphi \equiv \varphi \vee \mathcal{N}\phi$, while $\varphi_1\mathcal{S}\varphi_2\equiv \varphi_1\wedge \neg (\varphi_1\mathcal{R}\neg (\varphi_1\vee \varphi_2))$.

\section{Model checking algorithm}%
\label{sec:ModelChecking}

In this section we describe a model checking algorithm for SLSC, which is
an adaptation of the standard model checking algorithm for the \emph{Computational Tree Logic} (CTL)~\cite{baier2008principles}.
Given a finite simplicial model $\mathcal{M} = ( \mathcal{K}, P, \nu)$, a formula $\varphi$, and the used \emph{adjacency relation}, the proposed algorithm returns a set $\Sigma$ of simplices  satisfying $\varphi$ in $\mathcal{M}$.
The pseudocode of  function $\mathtt{Sat}$ is reported in Algorithm~\ref{alg:mc}. This function is inductively defined on the structure of $\varphi$ and computes the resulting set following a bottom-up approach.
When $\varphi$ is of the form $a, \top, \lnot \varphi_1, \varphi_1 \wedge \varphi_2$ the definition of $\mathtt{Sat}(\mathcal{M}, \varphi, \mathcal{C})$ is straightforward.
For this reason we discuss with more details the two spatial operators.

To compute the set of simplicial complexes satisfying $\mathcal{N}\varphi $, function  $\mathtt{Sat}$ first computes the set  $\Sigma$ of simplicial complexes satisfying $\varphi$. After that, the resulting set is computed by invoking function $\mathtt{Adj}(\Sigma, \mathcal{C})$ returning the set of simplicial complexes that are adjacent to the elements of $\Sigma$ according to $\mathcal{C}$:
\[
\mathtt{Adj}(\Sigma, \mathscr{C}) = \{ \sigma~|~\exists \sigma'. \sigma , \mathscr{C} \sigma'\}
\]

\begin{figure}
\centering
\includegraphics[scale=0.45]{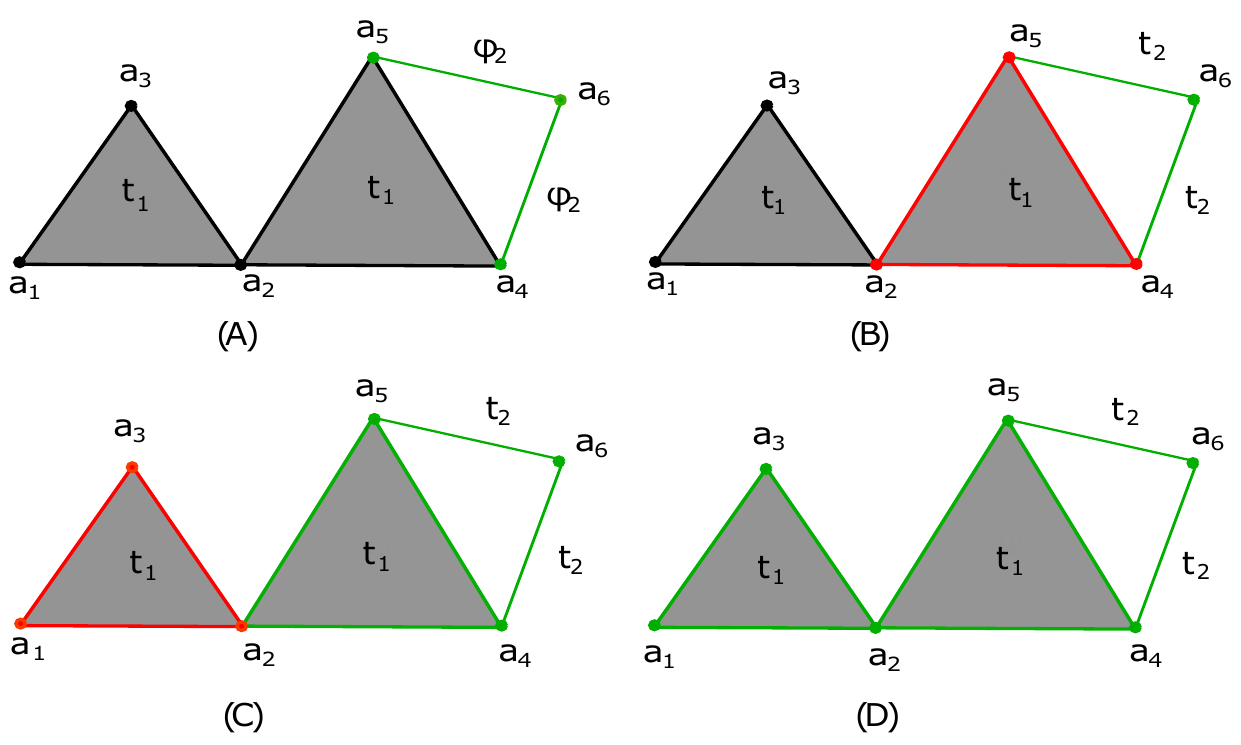}
\caption{Model checking $\varphi_1 \mathcal{R}\varphi_2$ using the lower adjacency.}%
\label{imm:fig5}
\end{figure}

When $\varphi$ is of the form $\varphi_1 \mathcal{R} \varphi_2$,  function $\mathtt{Sat}$ relies on the function $\mathtt{Reach}$ defined in Algorithm~\ref{alg:reach}.
Function $\mathtt{Reach}$ takesa simplicial model $\mathcal{M}$, two formulas $\varphi_1$ and $\varphi_2$, and an adjacency relation, $ \mathscr{C}$,  as parameters. The function returns the set of simplices in $\mathcal{K}$ that can \emph{reach} elements in $\Sigma_2$ by only traversing elements in $\Sigma_1$.
This set is computed iteratively via a flooding that starts from all the simplices in $\Sigma_2 =\mathtt{Sat}(\mathcal{M}, \varphi_2,\mathscr{C})$ and, at each step, add the adjacent simplices that are in $\Sigma_1 =\mathtt{Sat}(\mathcal{M}, \varphi_1,\mathscr{C})$. %
The evolution of this algorithm is illustrated in an informal way in Figure~\ref{imm:fig5} for the formula $ t_1 \mathcal{R} t_2$ using the lower adjacency and  considering the simplicial complex in Figure~\ref{imm:fig3}.
Firstly, all complexes that satisfy $t_2$ (the green simplices in Figure~\ref{imm:fig5}-(A)) are included to $T$. Thus, $T = \{[a_5,a_6], [a_6,a_4] \}$. 
In the second step, the algorithm selects all complexes that are lower adjacent to $[a_5,a_6]$ and $[a_6,a_4]$ and adds them to $T$.
The newly added simplices are shown in red in Figure~\ref{imm:fig5}-(B), while $[a_5,a_6]$ and $[a_6,a_4]$are illustrated in green. 
The algorithm terminates when there are not new simplices to add to the set $T$. %
In the example this happens after other two iterations as reported in Figure\ref{imm:fig5}-(C) and Figure\ref{imm:fig5}-(D). 

\begin{algorithm}[ht]
$\mathtt{Sat}(\mathcal{M}, \varphi, \mathscr{C})$ \\
\SetAlgoLined
\SetKwData{Left}{left}\SetKwData{This}{this}\SetKwData{Up}{up}
\SetKwFunction{Union}{Union}\SetKwFunction{FindCompress}{FindCompress}
\SetKwInOut{Input}{input}\SetKwInOut{Output}{output}
\Input{ Simplicial Model $\mathcal{M}=(\mathcal{K}, P, \nu)$, a formula $\varphi$, an adjacency relation $\mathscr{C}$}
\Output{Set of simplices $\Sigma = \{\sigma \in \mathcal{K} : \sigma \in \llbracket  \varphi  \rrbracket_{\mathscr{C}} \}$}
{\bf Match} $\varphi$ \\
\Switch{X}{
            \Case{$a$}{
            \Return{$\nu(a)$}
            }
            \Case{$\top$}{
            \Return{$\mathcal{K}$}
            }
            \Case{$  \lnot \varphi$}{
            $\Sigma = \mathtt{Sat}(\mathcal{M}, \varphi, \mathscr{C})$\\
            \Return{$\mathcal{K} \setminus \Sigma$}
            }
             \Case{$ \varphi_1 \wedge \varphi_2 $}{
             $\Sigma = \mathtt{Sat}(\mathcal{M}, \varphi_1, \mathscr{C})$\\
             $\Sigma = \mathtt{Sat}(\mathcal{M}, \varphi_2, \mathscr{C})$\\
             \Return{$\Sigma \cap \Sigma'$}
             }

              \Case{$ \mathcal{N} \varphi $}{
             $\Sigma = \mathtt{Sat}(\mathcal{M}, \varphi,\mathscr{C})$ \;
            \Return{$\mathtt{Adj}(\Sigma,\mathcal{C})$}
             }

             \Case{$\varphi_1 \mathcal{R} \varphi_2$}{
                       {\bf return} $\mathtt{Reach}(\mathcal{M}, \varphi_1, \varphi_2)$
             }
}{}

\caption{Decision procedure for the model checking problem}%
\label{alg:mc}
\end{algorithm}

\begin{algorithm}
\DontPrintSemicolon

$\mathtt{Reach}(\mathcal{M}, \Sigma_1, \Sigma_2, \mathscr{C})$ \\
\SetAlgoLined
\SetKwData{Left}{left}\SetKwData{This}{this}\SetKwData{Up}{up}
\SetKwFunction{Union}{Union}\SetKwFunction{FindCompress}{FindCompress}
\SetKwInOut{Input}{input}\SetKwInOut{Output}{output}
\Input{ Simplicial Model $\mathcal{M}$, sets of simplicial complexes $\Sigma_1$, $\Sigma_2$}
\Output{Set of simplices $\Sigma = \bigcup_{i=0}^{\infty} R^{i}_{\mathscr{C}} (\Sigma_1  , \Sigma_2)$}
$\Theta = \Sigma_2 $\\
$T = \Sigma_2 $\\
\While{$\Theta \neq \emptyset$}{
$\Theta '  := \emptyset$ \;
\For{$\sigma \in \Theta$}{
$\Sigma ' = \Sigma_1 \cap  \{\bar{\sigma} \in \mathcal{K}: \bar{\sigma} \mathscr{C} \sigma \} $\\
$\Theta '  = \Theta '  \cup (\Sigma '  \setminus T)$\\
$T = T \cup \Sigma'$
}
$\Theta = 	\Theta ' $
}
{\bf return} $T$
\caption{Checking \textit{reachability} formulas in a simplicial complex}

\label{alg:reach}
\end{algorithm}

In order to address termination, complexity and correctness of our algorithms, we define the notion of $\it{size}$ of a formula.
\begin{defi}[Size of a Formula] For any  SLSC formula $\varphi$, let $\mathtt{size}(\varphi)$ be inductively defined as follows:
\begin{itemize}
\item $\mathtt{size}(\top) = \mathtt{size}(p) = 1$
\item $\mathtt{size}(\lnot \varphi) = \mathtt{size}(\mathcal{N} \varphi) = 1+ \mathtt{size}(\varphi) $
\item $ \mathtt{size} (\varphi_1 \wedge \varphi_2) = \mathtt{size} (\varphi_1 \mathcal{R} \varphi_2)  =  \mathtt{size} (\varphi_1) + \mathtt{size} (\varphi_2)  $
\end{itemize}
\end{defi}

\noindent
The following theorem guarantees that function $\mathtt{Sat}$ terminates in a number of steps that is linear with the size of the model and with the size of the formula.

\begin{thm}
For any finite simplicial model $\mathcal{M} = (\mathcal{K}, P, \nu)$ and SLSC formula $\varphi$, $\mathtt{Sat}(\mathcal{M}, \varphi, \mathscr{C})$ terminates in $\mathcal{O}(\mathtt{size}(\varphi) \cdot |\mathcal{K}|)$ steps, where $|\mathcal{K}|$ denotes the number of simplicial complexes in $\mathcal{K}$.
\end{thm}

\begin{proof} The proof is by induction on the structure of formulas. \\
\textit{Base of Induction}. If $\varphi = \top$ or $\varphi =  a$, the statement follows directly from the definition of $\mathtt{Sat}$. Indeed, in both these cases the computation terminates in $1$ step and returns a set containing at most $|\mathcal{K}|$ elements.

\noindent
\textit{Inductive Hypothesis}.
Let $\varphi_1$ and $\varphi_2$ be two formulas such that, for any finite simplicial model $\mathcal{M}= (\mathcal{K}, P, \nu)$, function $\mathtt{Sat}(\mathcal{M}, \varphi_i, \mathscr{C})$  terminates in at most $\mathcal{O}(\mathtt{size}( \varphi_i) \cdot |\mathcal{K}|)$ steps.

\noindent
\textit{Inductive Step}. We can distinguish the following cases:
\begin{itemize}
\item[-] $\varphi = \lnot \varphi_1$. By inductive hypothesis,
we have that $\Sigma_1 = \mathtt{Sat}(\mathcal{M}, \varphi_1, \mathscr{C})$ is computed in $\mathcal{O}( \mathtt{size}(\varphi_1) \cdot |\mathcal{K}|)$. By definition of function $\mathtt{Sat}$,  $\mathtt{Sat}(\mathcal{M}, \lnot \varphi_1, \mathscr{C}) = \mathcal{K} \setminus \Sigma_1$. Since $\mathcal{K}$ is finite, we need at most $\mathcal{O}( |\mathcal{K}|)$ steps to compute $\mathcal{K} \setminus \Sigma_1$.
This means that the computation of $\mathtt{Sat}(\mathcal{M}, \lnot \varphi, \mathscr{C})$ terminates in at most  $\mathcal{O}(\mathtt{size}(\varphi_1) \cdot |\mathcal{K} |) + \mathcal{O}(|\mathcal{K}|)$. Hence:
\[\mathcal{O}(\mathtt{size}(\varphi_1) \cdot |\mathcal{K} |  ) + \mathcal{O}(|\mathcal{K} |) = \mathcal{O}((\mathtt{size}(\varphi_1) +1) \cdot |\mathcal{K} | )=  \mathcal{O}(\mathtt{size}(\lnot \varphi_1) \cdot |\mathcal{K} | )\]
\item[-] $\varphi = \varphi_1 \wedge \varphi_2$. In this case we have that $\mathtt{Sat}(\mathcal{M}, \varphi_1 \wedge \varphi_2, \mathscr{C}) = \mathtt{Sat}(\mathcal{M}, \varphi_1, \mathscr{C}) \ \cap \ \mathtt{Sat}(\mathcal{M}, \varphi_2, \mathscr{C})$. Moreover, by inductive hypothesis, we have that the computation of
$\mathtt{Sat}(\mathcal{M}, \varphi_i, \mathscr{C})$ terminates in $\mathcal{O}(\mathtt{size}(\varphi_i) \cdot|\mathcal{K}|) $ for all $i = 1,2$.

It is easy to observe that, since $\mathcal{K}$ is finite, the computation of $\Sigma_1 \cap \Sigma_2$ can be computed in at most $\mathcal{O}(|\mathcal{K}|)$. This means that $\mathtt{Sat}(\mathcal{M}, \varphi_1 \wedge \varphi_2, \mathscr{C})$ terminates in at most  $\mathcal{O}(\mathtt{size}(\varphi_1) \cdot |\mathcal{K} |) + \mathcal{O}(\mathtt{size}(\varphi_2) \cdot |\mathcal{K} |) + \mathcal{O}(|\mathcal{K}|)$. Hence:

\[\mathcal{O}(\mathtt{size}( \varphi_1) \cdot |\mathcal{K}|) +\mathcal{O}(\mathtt{size}( \varphi_2) \cdot |\mathcal{K}|) + \mathcal{O}( |\mathcal{K}|)  = \mathcal{O}((\mathtt{size}(\varphi_1) + \mathtt{size}( \varphi_2) +1) \cdot |\mathcal{K}|)\]
\[= \mathcal{O}((\mathtt{size}( \varphi_1) + \mathtt{size}( \varphi_2)) \cdot |\mathcal{K}|) = \mathcal{O}(\mathtt{size}( \varphi_1 \wedge \varphi_2 )\cdot |\mathcal{K}|)\]

\item[-] $\varphi =\mathcal{N}\varphi_1$. In this case we have that $ \mathtt{Sat}(\mathcal{M}, \mathcal{N} \varphi_1,\mathscr{C}) = \mathtt{Adj}(\Sigma_1)$ where $\Sigma_1 =  \mathtt{Sat}(\mathcal{M},\varphi_1,\mathscr{C})$. By inductive hypothesis, we have the computation of $\mathtt{Sat}(\mathcal{M},\varphi_1,\mathscr{C})$ terminates in $\mathcal{O}(\mathtt{size}(\varphi_1) \cdot|\mathcal{K}|)$. It is easy to see that, since $\mathcal{K}$ is finite, the computation of $\mathtt{Adj}(\Sigma_1)$ requires $\mathcal{O}( |\mathcal{K} |)$ steps. Hence, $\mathtt{Sat}(\mathcal{M}, \mathcal{N} \varphi, \mathscr{C})$ terminates in at most  $\mathcal{O}(\mathtt{size}(\varphi_1)\cdot |\mathcal{K} | ) + \mathcal{O}(|\mathcal{K}|)$. Therefore:
\[\mathcal{O}(\mathtt{size}(\varphi_1) \cdot |\mathcal{K} | ) + \mathcal{O}(|\mathcal{K}|) = \mathcal{O}((\mathtt{size}(\varphi_1) +1) \cdot |\mathcal{K}| ) = \mathcal{O}(\mathtt{size}(\mathcal{N} \varphi_1) \cdot |\mathcal{K} |)\]

\item[-] $\varphi = \varphi_1 \mathcal{R} \varphi_2$. We have that $\mathtt{Sat}({\mathcal{M}, \varphi_1 \mathcal{R} \varphi_2})$ invokes function $\mathtt{Reach}$ with parameters $\Sigma_1 = \mathtt{Sat}(\mathcal{M}, \varphi_1,\mathscr{C})$ and $\Sigma_2 = \mathtt{Sat}(\mathcal{M}, \varphi_2,\mathscr{C})$. Starting from the elements in $\Sigma_2$, the element in $\Sigma_1$ are added to the set $T$. We can observe that, in function $\mathtt{Reach}$, each simplicial complex in $\mathcal{K}$ is taken into account only one time.  This means that this function terminates after at most $\mathcal{O}(|\mathcal{K}|)$ steps.
By inductive hypothesis, we also have that the computations of $\Sigma_1$ and $\Sigma_2$ are computed in at most
 $\mathcal{O}(\mathtt{size}( \varphi_1) \cdot |\mathcal{K}| )$ and  $\mathcal{O}(\mathtt{size}( \varphi_2) \cdot |\mathcal{K}| )$ steps.

Summing up, the computation of $\mathtt{Sat}({\mathcal{M}, \varphi_1 \mathcal{R} \varphi_2})$ terminates in at most
 $\mathcal{O}(\mathtt{size}( \varphi_1) \cdot |\mathcal{K}|) + \mathcal{O}(\mathtt{size}( \varphi_2) \cdot |\mathcal{K} | ) + \mathcal{O}( |\mathcal{K}| ) = \mathcal{O}((\mathtt{size}( \varphi_1) +\mathtt{size}( \varphi_2) +1) \cdot |\mathcal{K}| ) = \mathcal{O}(\mathtt{size}( \varphi_1 \mathcal{R} \varphi_2) \cdot |\mathcal{K}| ) $
\qedhere
\end{itemize}
\end{proof}

\begin{thm}
Let $\mathscr{C}$ be an element of $\{ \smile, \frown, \multimapdotboth \}$. For any finite simplicial model $\mathcal{M} = (\mathcal{K}, P, \nu)$ and a formula $\varphi$, $\mathtt{Sat}(\mathcal{M}, \varphi, \mathscr{C})= \llbracket \varphi \rrbracket_{\mathscr{C}}$.
\end{thm}

\begin{proof}
The proof proceeds by induction on the syntax of the formulae.

\noindent
\textit{Base of Induction}. If  $\varphi = \top$ or $\varphi =a $ the statement follows directly from the definition of function $\mathtt{Sat}$ and from Definition~\ref{Semantics}.

\noindent
\textit{Inductive Hypothesis (IH)}. Let $\varphi_1$ and $\varphi_2$ be two formulas such that for any simplicial model  $\mathcal{M} = (\mathcal{K}, P, \nu)$, for any $\sigma \in \mathcal{M}$, $\sigma \in \mathtt{Sat}(\mathcal{M}, \varphi_i,\mathscr{C}), \ \forall i = \{1,2\} \iff \sigma_i \in \llbracket \varphi_i \rrbracket_{\mathscr{C}}, \   \forall i =\{1,2\} $

\medskip
\noindent
\textit{Inductive Step.} We can distiguish the following cases:
\begin{itemize}
\item[-] $\! \varphi = \! \lnot \varphi_1: \! \sigma \in \mathtt{Sat} (\mathcal{M}, \lnot \varphi_1,\mathscr{C})\!  \xLeftrightarrow[\text{}]{ \text{}\mathtt{Sat} }  \!  \sigma \in \ \mathcal{K} \setminus \mathtt{Sat} (\mathcal{M}, \varphi_1,\mathscr{C}) \xLeftrightarrow[\text{}]{\text{IH}}  \sigma \in \ \mathcal{K} \setminus   \llbracket \varphi_1 \rrbracket_{\mathscr{C}}  \xLeftrightarrow[\text{}]{\text{Def.\ref{Semantics}}}  \sigma \in \ \llbracket \lnot \varphi_1 \rrbracket_{\mathscr{C}}$

\item[-] $\varphi =  \varphi_1 \wedge  \varphi_2  : \sigma \in \mathtt{Sat} (\mathcal{M},  \varphi_1 \wedge  \varphi_2,\mathscr{C}) \xLeftrightarrow[\text{}]{\text{}\mathtt{Sat}}  \sigma \in  \mathtt{Sat} (\mathcal{M},  \varphi_1,\mathscr{C}) \ \cap  \    \mathtt{Sat} (\mathcal{M},  \varphi_2,\mathscr{C}) \xLeftrightarrow[\text{   }]{\text{    }}  \sigma \in  \mathtt{Sat} (\mathcal{M},  \varphi_1,\mathscr{C}) \ \mbox{ and } \  \sigma \in \mathtt{Sat} (\mathcal{M},  \varphi_2,\mathscr{C}) \xLeftrightarrow[\text{}]{\text{IH}}   \sigma \in \llbracket \varphi_1 \rrbracket_{\mathscr{C}} \cap   \llbracket \varphi_2  \rrbracket_{\mathscr{C}} \xLeftrightarrow[\text{}]{\text{Def.~\ref{Semantics}}}   \llbracket  \varphi_1 \wedge \varphi_2 \rrbracket_{\mathscr{C}} $
\item[-] $\varphi =  \mathcal{N}\varphi_1  : \sigma \in \mathtt{Sat} (\mathcal{M},  \mathcal{N}\varphi_1,\mathscr{C})   \xLeftrightarrow[\text{}]{\text{Fun.} \mathtt{Sat}} \sigma \in \ \mathtt{Adj}(\mathtt{Sat} (\mathcal{M}, \varphi_1,\mathscr{C}))
\xLeftrightarrow[\text{}]{ }  \sigma \in \ \{\sigma '  \in \mathcal{K}: \sigma '  \mathscr{C}\sigma_1$ $  \ \mbox{and} \  \sigma_1 \in \mathtt{Sat} (\mathcal{M}, \varphi_1,\mathscr{C}) \} \xLeftrightarrow[\text{}]{\text{IH}} \sigma \in \ \{ \sigma '  \in \mathcal{K}: \sigma_1 \in  \llbracket \varphi_1 \rrbracket_{\mathscr{C}}$
$\mbox{ and } \sigma ' \mathscr{C} \sigma_1 \} \xLeftrightarrow[\text{}]{\text{Def.~\ref{Semantics}}} \sigma \in \llbracket  \mathcal{N} \varphi_1 \rrbracket_{\mathscr{C}} $
\item[-]  $\varphi =  \varphi_1 \mathcal{R}\varphi_2$: we prove that $\sigma \in \mathtt{Reach}(\mathcal{M}, \varphi_1, \varphi_2)$ if and only if $\sigma \in \llbracket  \varphi_1 \mathcal{R}\varphi_2 \rrbracket_{\mathscr{C}}$.  The proof proceeds by induction on the structure of formulas.  Note that, by inductive hypothesis, we have that:
\[\sigma_i \in  \mathtt{Sat}(\mathcal{M}, \varphi_i,\mathscr{C}) \ \forall i = \{1,2\}  \iff \sigma_i \in \llbracket \varphi_i \rrbracket_{\mathscr{C}} \ \ \forall i =\{1,2\}  \]
We let $T_k$ denote the set of elements of variable $T$ at iteration $k$ in $\mathtt{Reach}$. Thus, we must prove that $\sigma \in T_k$ if and only if $\llbracket  \varphi_1 R^k \varphi_2 \rrbracket_{\mathscr{C}}$ for all $k$.  We proceed by induction on $k$.
\begin{itemize}
\item[-] Base of Induction, $k=0$: $\sigma \in T_0  \xLeftrightarrow[\text{}]{\text{ }\mathtt{Reach}} \sigma \in \mathtt{Sat}(\mathcal{M}, \varphi_2,\mathscr{C})\xLeftrightarrow[\text{}]{\text{IH}} \sigma \in \llbracket  \varphi_2 \rrbracket_{\mathscr{C}}  \xLeftrightarrow[\text{}]{\text{Def }\ref{Semantics}} \sigma \in \llbracket \varphi_1 R^0  \varphi_2 \rrbracket_{\mathscr{C}} $.

\item[-] Inductive Hypothesis, $\forall k \leq n$: we assume that $\sigma \in T_k$ if and only if $\sigma \in   \llbracket  \varphi_1 R^k \varphi_2 \rrbracket_{\mathscr{C}} $ and we prove that  $\sigma \in T_{n+1} \iff \sigma \in   \llbracket  \varphi_1 R^{n+1} \varphi_2 \rrbracket_{\mathscr{C}} $.

\begin{align*}
\sigma \in T_{n+1}
    &\xLeftrightarrow[\text{}]{\mathtt{Reach}} \sigma \in T_n \cup \ \Sigma'_{n+1} \\
    &\xLeftrightarrow[\text{}]{\mathtt{IH}} \sigma \in \{\bar{\Sigma} \in \mathcal{K}:  \bar{\sigma}  \in \llbracket \varphi_1 \rrbracket_{\mathscr{C}}, \exists \bar{\sigma}' \in \llbracket \varphi_1 R^{n-1} \varphi_2 \rrbracket_{\mathscr{C}} \mbox{ and } \bar{\sigma}  \mathscr{C}  \bar{\sigma'} \} \cup \Sigma'_{n+1} \\
    &\xLeftrightarrow[\text{}]{\mathtt{Reach}} \sigma \in \{\bar{\sigma} \in \mathcal{K}:  \bar{\sigma}  \in \llbracket \varphi_1 \rrbracket_{\mathscr{C}}, \exists \bar{\sigma}' \in \llbracket \varphi_1 R^{n-1} \varphi_2 \rrbracket_{\mathscr{C}} \mbox{ and } \bar{\sigma}  \mathscr{C}  \bar{\sigma'} \} \\
    & \hspace{2cm} {} \cup \{  \bar{\sigma''} \in \mathcal{K}: \bar{\sigma''} \mathscr{C} \bar{\sigma}  \} \cap \Sigma_1 \\
    &\xLeftrightarrow[\text{}]{} \sigma \in \{ \bar{\sigma''}, \bar{\sigma} \in \mathcal{K}: \bar{\sigma''},  \bar{\sigma} \in \llbracket \varphi_1 \rrbracket_{\mathscr{C}} \mbox{ and } \exists \bar{\sigma'} \in \llbracket \varphi_1 R^{n-1} \varphi_2 \rrbracket_{\mathscr{C}} \mbox{ and }  \bar{\sigma'}  \mathscr{C}  \bar{\sigma}  \mbox{ and }  \bar{\sigma''}  \mathscr{C}  \bar{\sigma} \}  \\
    &\xLeftrightarrow[\text{}]{\text{Def }\ref{Semantics}} \sigma \in \{\bar{\sigma''} \in \mathcal{K}: \bar{\sigma''}  \in  \llbracket \varphi_1 R^{n} \varphi_2 \rrbracket_{\mathscr{C}} \mbox{ and } \bar{\sigma''}  \mathscr{C}  \bar{\sigma} \} \\
    &\xLeftrightarrow[\text{}]{\text{Def }\ref{Semantics}} \sigma \in \llbracket \varphi_1 R^{n+1} \varphi_2 \rrbracket_{\mathscr{C}}
    \tag*{\qedhere}
\end{align*}
\end{itemize}
\end{itemize}
\end{proof}

\section{Expressive power of SLSC}%
\label{Bisimulations}

In this section, we introduce two equivalence relations among simplicial complexes that allow us to study the expressiveness of the proposed logic. The two equivalences are variants of strong bisimulation and branching bisimulation already defined in the literature~\cite{S2011, DV95}. Such equivalences, called $\mathscr{C}$-bisimulation and $\mathscr{C}$-branching bisimulation,  will be used to equate simplices that satisfy the same formulas. Two fragments of the proposed logic are identified that fully characterise the two proposed equivalences. Moreover, bisimulations between models based on simplicial complexes for epistemic logic have been proposed in~\cite{goubault2021simplicial,van2021dynamic}.
Both the proposed equivalences are parameterised with respect to the considered adjacency relation $\mathscr{C}$.

The first equivalence we consider is the $\mathscr{C}$-bisimulation. Following a standard approach, this equivalence identifies two simplicial complexes that are not be distinguished when one observes the adjacent complexes identified by the relation $\mathscr{C}$.

\begin{defi}[$\mathscr{C}$-bisimulation on simplicial models]%
\label{bisimulation}
Let $\mathcal{M}_1 = (\mathcal{K}_1, P, \nu_1)$ and $\mathcal{M}_2 = (\mathcal{K}_2, P, \nu_2)$ be two simplicial models. Let $\mathscr{C}$ be an element of $\{ \smile, \frown, \multimapdotboth \}$. A $\mathscr{C}$-bisimulation between $\mathcal{M}_1$ and  $\mathcal{M}_2$ is a non-empty binary relation $\mathcal{B}^{\mathscr{C}}\subseteq \mathcal{K}_1 \times \mathcal{K}_2$ such that  for any  $\sigma_1\in \mathcal{K}_1$ and $\sigma_2 \in \mathcal{K}_2$ whenever $(\sigma_1, \sigma_2) \in \mathcal{B}^{\mathscr{C}}$  we have that:
\begin{enumerate}
\item[a.]  $\nu_1(\sigma_1)=a$ if and only if $\nu_2(\sigma_2)=a$ for all $a \in P$ $\sigma_1 \equiv_{\nu} \sigma_2$;
\item[b.] for all $\sigma_1 '$ such that $\sigma_1  \mathscr{C} \sigma_1 '$, there exists $\sigma_2 '$ such that $\sigma_2  \mathscr{C} \sigma_2 '$ and $(\sigma_1 ' , \sigma_2 ' ) \in \mathcal{B}^{\mathscr{C}}   $;
\item[c.] for all $\sigma_2 '$ such that $\sigma_2  \mathscr{C} \sigma_2 '$, there exists $\sigma_1 '$ such that $\sigma_1  \mathscr{C} \sigma_1 ' $ and $(\sigma_1 ' , \sigma_2 ' ) \in \mathcal{B}^{\mathscr{C}}$.

\end{enumerate}%
\label{def:bisimulation}
\end{defi}

\begin{defi} Let $\mathcal{M}_1 = (\mathcal{K}_1, P, \nu_1)$ and  $\mathcal{M}_2=(\mathcal{K}_2, P, \nu_2) $ be two simplicial models. We say that $\sigma_1\in \mathcal{K}_1$ and $\sigma_2\in \mathcal{K}_2$ are \emph{$\mathscr{C}$-bisimilar}, write $\sigma_1\leftrightarroweq^{\mathscr{C}} \sigma_2$, whenever there exists a $\mathscr{C}$-bisimulation $\mathcal{B}^{\mathscr{C}}$ between $\mathcal{M}_1$ and  $\mathcal{M}_2$ such that $(\sigma_1,  \sigma_2) \in \mathcal{B}^{\mathscr{C}}$.

Moreover, we will say that $\mathcal{M}_1$ and $\mathcal{M}_2$ are \emph{$\mathscr{C}$-bisimilar}, written $\mathcal{M}_1 \leftrightarroweq^{\mathscr{C}}\mathcal{M}_2$, if and only if:
\begin{itemize}
\item  for all $\sigma_1 \in \mathcal{K}_1$, there is $ \sigma_2 \in \mathcal{K}_2$ such that $\sigma_1 \leftrightarroweq^{\mathscr{C}} \sigma_2$;
\item  for all $\sigma_2 \in \mathcal{K}_2$, there is $ \sigma_1 \in \mathcal{K}_1$ such that $\sigma_1 \leftrightarroweq^{\mathscr{C}} \sigma_2$.
\end{itemize}
\end{defi}

 \begin{figure}[tbp]
\centering
\includegraphics[scale=0.55]{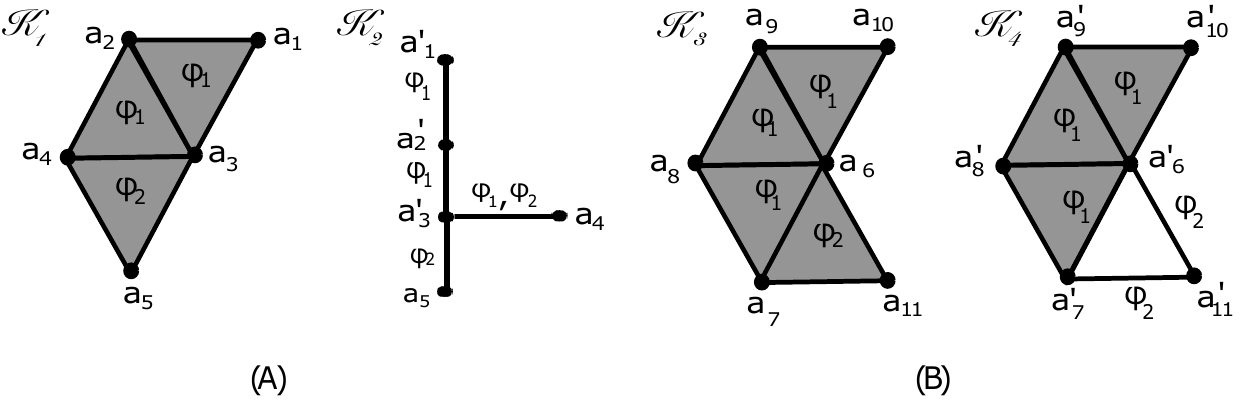}
\caption{Examples of bisimilar ($\mathcal{K}_1$ and $\mathcal{K}_2$) and non-bisimilar  ($\mathcal{K} _3$ and $\mathcal{K}_4$) simplicial models.}%
\label{imm:bisimulation}
\end{figure}

\noindent
Consider the two models $\mathcal{M}_1$ and $\mathcal{M}_2$, whose simplicial complexes are $\mathcal{K}_1$ and $\mathcal{K}_2$ illustrated in Figure~\ref{imm:bisimulation}-(A). 
$\mathcal{M}_1$ is lower-bisimilar  ($\leftrightarroweq^{\mathscr{\smile}}$) to $\mathcal{M}_2$ because 
each simplex in $\mathcal{K}_1$ corresponds to another simplex in $\mathcal{K}_2$ and both satisfy the  same formula. In fact, take the pair $([a_1,a_2,a_3], [a_1', a_2 ']) \in \mathcal{B}^{\smile} $: we need to show it obeys the hereditary conditions of Definition~\ref{bisimulation}.
$[a_1,a_2,a_3] \smile [ a_2,a_3,a_4], $ and $ [a_2,a_3,a_4]  \smile [a_3,a_4,a_5]$; however, $[a_1' ,a_2 '] \smile [a_2', a_3 ']$ and $ [a_2',a_3'] \smile [a_3',a_4']$.

Let us take into account the models $\mathcal{M}_3$ and $\mathcal{M}_4$ induced by the simplicial complexes $\mathcal{K}_3$ and $\mathcal{K}_4$ in Figure~\ref{imm:bisimulation}-(B). 
In this case we have that $\mathcal{M}_3$ is not lower-bisimilar to $\mathcal{M}_4$. Indeed, it is easy to see that no simplicial complex in $\mathcal{K}_4$ is lower-bisimilar to the simplicial  $[a_6, a_7, a_{11}]$ in $\mathcal{K}_3$.
The proposed spatial-bisimilarity can be used in the scenarios of Section~\ref{MotivatingExample} to identify symmetries in the spatial models. In particular, in the context of the \emph{scientific collaborations}, our bisimilarity identifies patterns of interactions among groups of researchers, regardless of their size.
Moreover, our equivalence is also able to detect areas having the same security level in the emergency rescue scenario.

Now, we are ready to establish the relationships between $\mathscr{C}$-bisimulation equivalence and the equivalence induced by the standard boolean operators equipped with the neighborhood operator of our logic. Given a logic language $L$ and an associated satisfaction relation interpreted over a model, the equivalence  $\sim^{\mathscr{C}}_{L}$, induced by $L$-formula, is given by
\[\sigma_1\sim^{\mathscr{C}}_{L}\sigma_2  \ \mbox{if and only if } (\forall \varphi \in L :  \sigma_1 \in   \llbracket  \varphi  \rrbracket_\mathscr{C} \iff   \sigma_2 \in   \llbracket  \varphi  \rrbracket_\mathscr{C}) \ .\]

\begin{defi}Let $\mathcal{M} =(\mathcal{K}, P, \nu)$ be a simplicial model. The syntax of an $L_N$ is defined inductively by the following grammar
$ \varphi ::= a \mid \top \mid \lnot \varphi_1 \mid \varphi_1 \wedge \varphi_2  \mid  \mathcal{N}  \varphi_2 $.
\end{defi}

However, to guarantee the intended equivalence we have to limit our attention to models with $\mathscr{C}$-\emph{bounded adjacency}. These are the class of models where each \emph{simplicial complex} is \emph{adjacent to} a finite number of other complexes.
This notion is reminiscent of the standard \emph{image-finiteness} used in the context of transition systems~\cite{S2011}.

\begin{defi}[$\mathscr{C}$-Bounded adjacency] Let $\mathcal{M} = (\mathcal{K}, P, \nu)$ be a simplicial model. We say that $\mathcal{M}$ has $\mathscr{C}$-\emph{bounded adjacency} if, for all $\sigma \in \mathcal{K}$, the cardinality of the set of adjacent simplices is finite.
\end{defi}

The following theorem guarantees that for any model $\mathcal{M}$ with $\mathscr{C}$-\emph{bounded adjacency},
$\mathscr{C}$-bisimulation equates simplicial complexes that satisfy the same formulas in $L_N$.

\begin{thm}%
\label{thm10}
Let $\mathcal{M} =(\mathcal{K}, P, \nu)$ be a simplicial model with $\mathscr{C}$-\emph{bounded adjacency},  and let $\mathscr{C}$ be an element of $\{ \smile, \frown, \multimapdotboth \}$.  $L_N$ and  $\leftrightarroweq^{\mathscr{C}} $ induce the same identification on simplicial complexes. Then, for all $\sigma_1$ and $\sigma_2$ in $\mathcal{K}$, $ \sigma_1 \leftrightarroweq^{\mathscr{C}} \sigma_2 \ \ \mbox{if and only if} \ \ \sigma_1 \sim^{\mathscr{C}}_{L_N} \sigma_2 \ . $
 \end{thm}
\begin{proof} ($``\Rightarrow"$) Suppose $ \sigma_1 \leftrightarroweq^{\mathscr{C}} \sigma_2$ and $\varphi \in L_N$. We prove that $\sigma_1 \in \llbracket \varphi \rrbracket_\mathscr{C}$ if and only if $\sigma_2 \in \llbracket \varphi \rrbracket_\mathscr{C}$  by induction on the syntax of $L_N$-formulae.\\
\noindent
\textit{Base of Induction.} If  $\varphi = \top$ or $\varphi = a$, then obviously $ \sigma_1 \in  \llbracket \varphi \rrbracket_\mathscr{C}$ and $ \sigma_2 \in  \llbracket \varphi \rrbracket_\mathscr{C}$.

\noindent
\textit{Inductive Hypothesis.} Let $\varphi_1$ and $\varphi_2$ be two formulas such that for any simplicial model$ \ \mathcal{M}$,  $\sigma_1 \in \llbracket \varphi_i \rrbracket_\mathscr{C} \ \mbox{ if  and only if} \  \sigma_2 \in \llbracket \varphi_i \rrbracket_\mathscr{C} \  \forall i =\{1,2\} $. \\
\textit{Inductive Step.}
\begin{enumerate}
\item If $\varphi = \lnot \varphi_1$, then, by definition $ \sigma_1 \in \llbracket \varphi \rrbracket_\mathscr{C}$ iff $ \sigma_1 \notin \llbracket \varphi_1 \rrbracket_\mathscr{C}$. By induction $ \sigma_1 \notin \llbracket \varphi_1 \rrbracket_\mathscr{C}$ iff $ \sigma_2 \notin \llbracket \varphi_1 \rrbracket_\mathscr{C}$. Again by definition $ \sigma_2 \notin \llbracket \varphi_1 \rrbracket_\mathscr{C}$ iff $ \sigma_2 \in \llbracket \varphi \rrbracket_\mathscr{C}$
\item If $\varphi = \varphi_1 \wedge \varphi_2$, then, by definition $\sigma_1 \in \llbracket \varphi \rrbracket_\mathscr{C}$ iff $ \sigma_1 \in \llbracket \varphi_1 \rrbracket_\mathscr{C}$ and $ \sigma_1 \in \llbracket \varphi_2 \rrbracket_\mathscr{C}$. By induction $ \sigma_1 \in \llbracket \varphi_1 \rrbracket_\mathscr{C}$ and $ \sigma_1 \in \llbracket \varphi_2 \rrbracket_\mathscr{C}$  iff $ \sigma_2 \in \llbracket \varphi_1 \rrbracket_\mathscr{C}$ and $ \sigma_2 \in \llbracket \varphi_2 \rrbracket_\mathscr{C}$. Again by definition $ \sigma_2 \in \llbracket \varphi \rrbracket_\mathscr{C}$.
\item
If $\varphi = \mathcal{N} \varphi_1$, then, by definition $\sigma_1 \in \llbracket \mathcal{N} \varphi_1 \rrbracket_\mathscr{C}$. Therefore, there exists $\sigma_1 ' \in  \mathcal{K}$,  such that
$\sigma_1 \mathscr{C} \sigma_1 ' \mbox { and } \sigma_1 ' \in \llbracket \varphi_1 \rrbracket_\mathscr{C} $.
By inductive hypothesis, $\sigma_1 \leftrightarroweq^{\mathscr{C}} \sigma_2$, there exists $\sigma_2' \in \mathcal{K}$ such that $ \sigma_2 \mathscr{C} \sigma_2 '  \mbox{ and } \sigma_1 ' \leftrightarroweq^{\mathscr{C}} \sigma_2 '$.
 By the inductive hypothesis, $\sigma_2 '  \in \llbracket \varphi_1 \rrbracket_\mathscr{C}$.
 Therefore, there exists $\sigma_2' \in \mathcal{K}$ such that $ \sigma_2 \mathscr{C} \sigma_2 '  \mbox{ and } \sigma_2 '  \in \llbracket \varphi_1 \rrbracket_\mathscr{C}. $
Hence, by definition, $\sigma_2 \in  \llbracket \mathcal{N} \varphi_1 \rrbracket_\mathscr{C}$.

\end{enumerate}
\noindent
($``\Leftarrow"$) The statement follows by proving that $ \sim^{\mathscr{C}}_{L_N} $ is a bisimulation. According to Definition~\ref{def:bisimulation}, we have to prove that for any $\sigma_1$ and $\sigma_2$ such that $\sigma_1 \sim^{\mathscr{C}}_{L_N} \sigma_2$ we have that:

\begin{enumerate}
\item[a.]   $\sigma_1 \in\nu_1(a)$ if and only if $\sigma_2 \in \nu_2(a)$ for all $a \in P$;
\item[b.] $\sigma_1 ' : \sigma_1  \mathscr{C} \sigma_1 ', \mbox{ there is } \sigma_2 ' : \sigma_2  \mathscr{C} \sigma_2 '$ and $\sigma_1 ' \sim^{\mathscr{C}}_{L_N} \sigma_2 '$;
\item[c.] $ \sigma_2 ' : \sigma_2  \mathscr{C} \sigma_2 '  , \mbox{ there is } \sigma_1 ' : \sigma_1  \mathscr{C} \sigma_1 ' $ and $\sigma_1 '\sim^{\mathscr{C}}_{L_N} \sigma_2 '$
\end{enumerate}

\noindent
We can observe that the first case is trivial, since $\sigma_1$ and $\sigma_2$ satisfy the same set of formulas.
We prove case $(b)$ by contradiction. Let us assume that
that there exists $\sigma_1' \in \mathcal{K}$ such that $\sigma_1\mathscr{C} \sigma_1'$ and, for any $\sigma_2'$ that is $\mathscr{C}$-adjacent to $\sigma_2$, $\sigma_1'\not \sim^{\mathscr{C}}_{L_N} \sigma_2'$.
This means that for any $\sigma_2'$ such that $\sigma_2\mathscr{C} \sigma_2'$ exists a formula $\varphi \in L$ such that $\sigma_1' \in \llbracket \varphi \rrbracket_\mathscr{C}$ and  $\sigma_2' \notin \llbracket \varphi \rrbracket_\mathscr{C}$.
Let us consider $\mathcal{U} = \{\sigma_2' | \sigma_2 \mathscr{C} \sigma_2'\}$.
We can assume $\mathcal{U}\not=\emptyset$. Indeed, in this case $\sigma_1\in \llbracket \mathcal{N}\top \rrbracket_\mathscr{C}$ while $\sigma_2\not\in \llbracket \mathcal{N}\top \rrbracket_\mathscr{C}$ that contradicts the hypothesis $\sigma_1 \sim^{\mathscr{C}}_{L_N} \sigma_2$.
Since $\mathcal{M}$ has $\mathscr{C}$-bounded adjacency, we have that $\mathcal{U}=\{\sigma_2^1,\ldots, \sigma_2^k\}$.
Hence, for any $i$ there exists $\varphi_i$ such that
$\sigma_1'\in \llbracket \varphi_i \rrbracket_\mathscr{C}$ while $\sigma_2^i\not\in \llbracket \varphi_i \rrbracket_\mathscr{C}$
We can now consider the formula $\varphi = \varphi_1 \wedge  \varphi_2 \wedge \dots \wedge \varphi_n, \sigma_1 ' \in \llbracket \varphi \rrbracket_\mathscr{C} \ \ \ \mbox{and } 
\sigma_2^{i} \not \in \llbracket \varphi \rrbracket_\mathscr{C} \mbox{ for every } i \in \{1, \dots, k\}$.

Hence, $\sigma_1 \in \llbracket \mathcal{N} \varphi \rrbracket_\mathscr{C}$, \ \mbox{while} \  $\sigma_2 \notin \llbracket \mathcal{N} (\varphi)  \rrbracket_\mathscr{C}$.
However, this contradicts the hypothesis that $\sigma_1 \sim^{\mathscr{C}}_{L_N} \sigma_2$.
The proof for case $c$.\ follows in a symmetric way.
\qedhere
\end{proof}

\begin{figure}[tbp]
\centering
\includegraphics[scale=0.75]{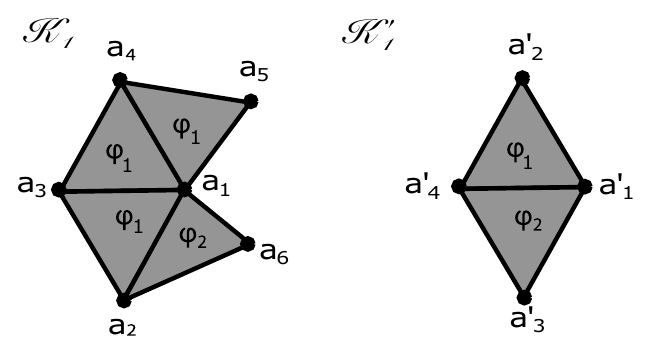}
\caption{An example of branching bisimilar simplicial models.}%
\label{imm:branchingbisimulation}
\end{figure}

The $\mathscr{C}$-bisimulation introduced above it is often too strong and discriminates elements  that could be considered equivalent. Let us consider, for instance, the two models in Figure~\ref{imm:branchingbisimulation}. The simplicial complexes $[a_1, a_4, a_5]$, $[a_1, a_3, a_4]$ and $[a_1, a_2, a_3]$ on the left part are distinguished by $\mathscr{C}$-bisimulation from $[a_1', a_2', a_4']$ on the right.
However, if we \emph{merge} the areas labelled with $\phi_1$, and ignoring the boundaries of the single elements, we can observe that the two models are in fact the same.

For this reason, in what follows, we will introduce the \textit{$\mathscr{C}$-branching bisimulation} relation over simplicial models.
This is an equivalence relationship that identifies simplices by considering a \emph{weaker} form of adjacency where an observer is only partially able to distinguish two adjacent simplicial complexes placed in similar context.

\begin{defi}[$\mathscr{C}^*$ adjacency relation]
Let $\mathcal{M} = (\mathcal{K}, P, \nu)$ be a simplicial model and $\mathscr{C}$ be an element of $\{ \smile, \frown, \multimapdotboth \}$, we let $\mathscr{C}^*$ denote the adjacency relation such that:
\begin{itemize}
    \item $\sigma \mathscr{C}^* \sigma$, for each $\sigma\in \mathcal{K}$;
    \item $\sigma_1 \mathscr{C}^* \sigma_2$, if and only if there exists $\sigma_1'\in \mathcal{K}$ such that $\sigma_1\equiv_{\nu} \sigma_1'$, $\sigma_1\mathscr{C} \sigma_1'$ and $\sigma_1' \mathscr{C}^* \sigma_2$.
\end{itemize}
\end{defi}

\begin{defi}[$\mathscr{C}$-branching bisimulation on simplicial models]
Let $\mathcal{M}_1 = (\mathcal{K}_1, P_1, \nu_1)$ and $\mathcal{M}_2 = (\mathcal{K}_2, P_2, \nu_2)$ be two simplicial models. Let $\mathscr{C}$ be an element of $\{ \smile, \frown, \multimapdotboth \}$. A $\mathscr{C}$-branching bisimulation between $\mathcal{M}_1$ and  $\mathcal{M}_2$ is a non-empty binary relation $\mathcal{E}^{\mathscr{C}}$ between their domains (that is, $\mathcal{E}^{\mathscr{C}} \subseteq \mathcal{K}_1 \times \mathcal{K}_2$)  such that for any $\sigma_1 \in \mathcal{K}_1$ and $ \sigma_2 \in \mathcal{K}_2$ whenever $(\sigma_1,\sigma_2) \in \mathcal{E}^{\mathscr{C}} $  we have that:
\begin{enumerate}
\item[a.] $\sigma_1 \equiv_{\nu} \sigma_2$;
\item[b.] if there exists $\ \sigma_1 '$ such that $\sigma_1 \mathscr{C} \sigma_1 '$, then either $(\sigma_1 ', \sigma_2) \in \mathcal{E}^{\mathscr{C}} $ or there exist $\sigma_2'$ and $\sigma_2''$ such that $\sigma_2 \mathscr{C}^{*} \sigma_2''\mathscr{C} \sigma_2'$, $(\sigma_1, \sigma_2'') \in \mathcal{E}^{\mathscr{C}} $ and $(\sigma_1', \sigma_2') \in \mathcal{E}^{\mathscr{C}} $;
\item[c.] if there exists $\ \sigma_2' $ such that $\sigma_2 \mathscr{C} \sigma_2 '$, then either $\sigma_1 \mathcal{E}^{\mathscr{C}} \sigma_2'$ or there exist $ \sigma_1'$ and $\sigma_1''$ such that $\sigma_1 \mathscr{C}^{*} \sigma_1''\mathscr{C} \sigma_1'$, $(\sigma_1'', \sigma_2) \in \mathcal{E}^{\mathscr{C}} $ and $(\sigma_1', \sigma_2') \in \mathcal{E}^{\mathscr{C}} $.
\end{enumerate}
\end{defi}

\begin{defi} Let $\mathcal{M}_1 = (\mathcal{K}_1, P, \nu_1)$ and  $\mathcal{M}_2=(\mathcal{K}_2, P, \nu_2) $ be two  simplicial models. We say that $\sigma_1\in \mathcal{K}_1$ and $\sigma_2\in \mathcal{K}_2$ are \emph{$\mathscr{C}$-branching bisimilar}, write $\sigma_1\leftrightarroweq_{br}^{\mathscr{C}} \sigma_2$,  whenever there exists a $\mathscr{C}$-branching bisimulation $\mathcal{E}^{\mathscr{C}}$ between $\mathcal{M}_1$ and  $\mathcal{M}_2$ such that $(\sigma_1, \sigma_2)\in  \mathcal{E}^{\mathscr{C}}$.

Moreover, we will say that $\mathcal{M}_1$ and $\mathcal{M}_2$ are \emph{$\mathscr{C}$-branching bisimilar}, written $\mathcal{M}_1 \leftrightarroweq_{br}^{\mathscr{C}}\mathcal{M}_2$, if and only if:
\begin{itemize}
\item for all $\sigma_1 \in \mathcal{K}_1$, exists $ \sigma_2 \in \mathcal{K}_2$ such that $\sigma_1\leftrightarroweq_{br}^{\mathscr{C}} \sigma_2$;
\item for all $\sigma_2 \in \mathcal{K}_2$, exists $ \sigma_1 \in \mathcal{K}_1$ such that $\sigma_1\leftrightarroweq_{br}^{\mathscr{C}} \sigma_2$.
\end{itemize}%
\label{branching}
\end{defi}


\noindent
Consider two models $\mathcal{M}_1$ and $\mathcal{M}_2$, whose simplicial complexes are $\mathcal{K}_1$ and $\mathcal{K}_2$, illustrated in Figure~\ref{imm:branchingbisimulation}. The two models are lower-branching bisimilar since they both have the same branching structure.
In Example~\ref{ex:simplicial_definitions:sc}, the branching bisimulation equates simplices, namely groups of coauthors, with the same chains of collaborations. Similarly, in Example~\ref{ex:simplicial_definitions:ee},  the branching bisimulation  guarantees that emergency rescues can reach the victim by passing through areas with the same security level without considering the exact number of steps.

\begin{lem}[Stuttering Lemma] Let $\mathcal{M} = (\mathcal{K}, P, \nu)$ be a simplicial model and $\mathscr{C}$ be an element of $\{ \smile, \frown, \multimapdotboth \}$. Let $\sigma_0,\sigma_1, \sigma_2, \dots, \sigma_n$ be simplices in $\mathcal{K}$ such that, for any $i<n$,  $\sigma_i \mathscr{C} \sigma_{i+1}$ and $\sigma_i\equiv_{\nu} \sigma_{i+1}$. If $\sigma_0 \leftrightarroweq_{br}^{\mathscr{C}} \sigma_n $, then $ \sigma_0 \leftrightarroweq_{br}^{\mathscr{C}} \sigma_i $ for all $0<i<n$.%
\label{LemmaStuttering}
\end{lem}

\begin{proof}  Suppose $\{\sigma_i \}_i$ as in the claim, and take
\[ \mathcal{R} = \{ (\sigma_0, \sigma_i): 0 < i  \leq n\} \ \cup \leftrightarroweq_{br}^{\mathscr{C}} \ . \]

We show that $ \mathcal{R}$ is a branching bisimulation. We have to prove that for any $(\hat{\sigma}_1,\hat{\sigma}_2)\in \mathcal{R}$
\begin{itemize}
    \item[a.] $\hat{\sigma}_1 \equiv_{\nu} \hat{\sigma}_2$
    \item[b.] if there exists $ \hat{\sigma_1 '} $ such that $\hat{\sigma_1} \mathscr{C} \hat{\sigma_1} '$, then either $(\hat{\sigma_1 '}, \hat{\sigma_2}) \in \mathcal{E}^{\mathscr{C}} $ or there exist $ \hat{\sigma_2'}$ and $\hat{\sigma_2''}$ such that $\hat{\sigma_2} \mathscr{C}^{*} \hat{\sigma_2''} \mathscr{C} \hat{\sigma_2'}$, $(\hat{\sigma_1}, \hat{\sigma_2''}) \in \mathcal{E}^{\mathscr{C}} $ and $(\hat{\sigma_1'}, \hat{\sigma_2'}) \in \mathcal{E}^{\mathscr{C}} $;
    \item[c.] if there exists $\ \hat{\sigma_2'} $ such that $\hat{\sigma_2} \mathscr{C} \hat{\sigma_2 '} $, then either $\hat{\sigma_1} \mathcal{E}^{\mathscr{C}} \hat{\sigma_2'}$ or there exist $\hat{\sigma_1'}$ and $\hat{\sigma_1''}$ such that $\hat{ \sigma_1} \mathscr{C}^{*} \hat{\sigma_1''} \mathscr{C} \hat{\sigma_1'}$, $(\hat{\sigma_1''}, \hat{\sigma_2}) \in \mathcal{E}^{\mathscr{C}} $ and $(\hat{\sigma_1'}, \hat{\sigma_2'}) \in \mathcal{E}^{\mathscr{C}} $.
\end{itemize}

\noindent
We can observe that if $(\hat{\sigma}_1,\hat{\sigma}_2)\in \leftrightarroweq_{br}^{\mathscr{C}}\subseteq \mathcal{R}$, all the properties above are satisfied by definition.  For this reason we consider pairs of the form $(\sigma_0,\sigma_i)$.
In this case the statement $(a)$ follows directly by assumption. To prove the statement $b$., let us consider consider a $\sigma_0'$ such that $\sigma_0 \mathscr{C} \sigma_0'$. By assumption, $\sigma_0 \leftrightarroweq_{br}^{\mathscr{C}} \sigma_n$, this implies that
either $(i)$ $\sigma_0 ' \leftrightarroweq_{br}^{\mathscr{C}} \sigma_n$, or  $(ii)$ there exist $\sigma_n ''$,  $\sigma_n '$ such that $\sigma_n \mathscr{C}^{*}\sigma_n ''\mathscr{C} \sigma_n '$, $\sigma_0 \leftrightarroweq_{br}^{\mathscr{C}} \sigma_n ''$ and $\sigma_0' \leftrightarroweq_{br}^{\mathscr{C}} \sigma_n '$.
If $(i)$ holds, the statement $(b)$ follows by observing that $\sigma_i \mathscr{C}^{*}\sigma_{n-1} \mathscr{C} \sigma_{n}$ and by observing that  $(\sigma_0, \sigma_{n-1}) \in \mathcal{R}$ and $(\sigma_0 ' , \sigma_{n}) \in \leftrightarroweq_{br}^{\mathscr{C}} \subseteq \mathcal{R}$.
If $(ii)$ holds, we can notice that $\sigma_i \mathscr{C}^{*} \sigma_n \mathscr{C}^{*} \sigma_n ''\mathscr{C}\sigma_n ' $. Hence, $\sigma_i \mathscr{C}^{*} \sigma_n ''\mathscr{C}\sigma_n '$ and the statement follows from the fact that $\sigma_0 \leftrightarroweq_{br}^{\mathscr{C}} \sigma_n$.
The statement $c$.\ follows easily by observing that $\sigma_{0}\mathscr{C}^{*} \sigma_{i}$. Hence, if $\sigma_i \mathscr{C} \sigma_i'$ we have that $\sigma_{0}\mathscr{C}^{*} \sigma_{i}\mathscr{C} \sigma_i'$ and, from the fact that $\leftrightarroweq_{br}^{\mathscr{C}}$ is reflexive, $\sigma_{i} \leftrightarroweq_{br}^{\mathscr{C}} \sigma_i$ and $\sigma_i' \leftrightarroweq_{br}^{\mathscr{C}} \sigma_i'$.
\qedhere
\end{proof}

\begin{defi}
Let $\mathcal{M} =(\mathcal{K}, P, \nu)$ be a simplicial model.
The syntax of $L_R$ is defined inductively by the following grammar
$ \varphi ::= a \mid \top \mid \lnot \varphi_1 \mid \varphi_1 \wedge \varphi_2  \mid \varphi_1 \mathcal{R}  \varphi_2 $.
\end{defi}

We are going to prove that formulas in $L_R$ have the same expressive power of  branching bisimulation. However, in order to obtain this result, we have to guarantee that, starting from a simplicial complex $\sigma$, a finite number of configurations satisfying a given set of atomic propositions can be reached. From a spatial point of view, this means that in the space, starting from a given area, we can reach a finite number of areas identified by a set of atomic propositions.

\begin{defi}%
\label{def:boundedreachability}
Let $\mathcal{M} =(\mathcal{K}, P, \nu)$ be a simplicial model. We say that $\mathcal{M}$ has \emph{bounded reachability} if and only if for any $\sigma\in \mathcal{K}$ and, for any set of atomic propositions $P'\subseteq P$, $\{ \sigma'| \lambda(\sigma')=P'\wedge \exists \sigma'': \sigma \mathscr{C}^{*} \sigma'' \mathscr{C} \sigma'\}$ is finite.
\end{defi}

\begin{thm}\label{thm11}
Let $\mathcal{M} =(\mathcal{K}, P, \nu)$ be a simplicial model with bounded reachability and let $\mathscr{C}$ be an element of $\{ \smile, \frown, \multimapdotboth \}$.  $L_R$ and  $\leftrightarroweq_{\mathscr{C}} $ induce the same identification on simplicial complexes. Then, for all $\sigma_1$ and $\sigma_2$ in $\mathcal{K}$:
$ \sigma_1 \leftrightarroweq_{br}^{\mathscr{C}} \sigma_2 \ \ \mbox{if and only if} \ \ \sigma_1 \sim^{\mathscr{C}}_{L_R} \sigma_2 \ . $
\end{thm}

\begin{proof} ($``\Rightarrow"$) Suppose $ \sigma_1 \leftrightarroweq_{br}^{\mathscr{C}} \sigma_2$ and $\varphi \in L_R$. We prove that $\sigma_1 \in \llbracket \varphi \rrbracket$ if and only if $\sigma_2 \in \llbracket \varphi \rrbracket$  by induction on the syntax of $L_R$-formulae.\\
\noindent
\textit{Base of Induction.} If  $\varphi = \top$, then obviously $ \sigma_1 \in  \llbracket \varphi \rrbracket_\mathscr{C}$ and $ \sigma_2 \in  \llbracket \varphi \rrbracket_\mathscr{C}$. \\
\textit{Inductive Hypothesis.} Let $\varphi_1$ and $\varphi_2$ be two functions such that for any simplicial model$ \ \mathcal{M}$,  $\sigma_1 \in \llbracket \varphi_i \rrbracket_\mathscr{C} \ \mbox{ if  and only if} \  \sigma_2 \in \llbracket \varphi_i \rrbracket_\mathscr{C} \  \forall i =\{1,2\} $. \\
\textit{Inductive Step.}
\begin{enumerate}
\item If $\varphi = \lnot \varphi_1$, then, by definition $ \sigma_1 \in \llbracket \varphi \rrbracket_\mathscr{C}$ iff $ \sigma_1 \notin \llbracket \varphi_1 \rrbracket_\mathscr{C}$. By induction $ \sigma_1 \notin \llbracket \varphi_1 \rrbracket_\mathscr{C}$ iff $ \sigma_2 \notin \llbracket \varphi_1 \rrbracket_\mathscr{C}$. Again by definition $ \sigma_2 \notin \llbracket \varphi_1 \rrbracket_\mathscr{C} $ iff $ \sigma_2 \in \llbracket \varphi \rrbracket_\mathscr{C}$

\item If $\varphi = \varphi_1 \wedge \varphi_2$, then by definition we have $\sigma_1 \in \llbracket \varphi \rrbracket_\mathscr{C} $ iff $ \sigma_1 \in \llbracket \varphi_1 \rrbracket_\mathscr{C} $ and $ \sigma_1 \in \llbracket \varphi_2 \rrbracket_\mathscr{C}$. By induction $ \sigma_1 \in \llbracket \varphi_1 \rrbracket_\mathscr{C} $ and $ \sigma_1 \in \llbracket \varphi_2 \rrbracket_\mathscr{C} $  iff $ \sigma_2 \in \llbracket \varphi_1 \rrbracket_\mathscr{C} $ and $ \sigma_2 \in \llbracket \varphi_2 \rrbracket_\mathscr{C} $. Again by definition $ \sigma_2 \in \llbracket \varphi \rrbracket_\mathscr{C} $.

\item If $\varphi = \varphi_1 \mathcal{R} \varphi_2$, suppose that $\sigma_1 \in \llbracket \varphi \rrbracket_\mathscr{C} $. We will prove that $\sigma_2 \in \llbracket \varphi \rrbracket_\mathscr{C} $. The reverse implication then follows by symmetry. We have to distinguish two cases:

\begin{enumerate}
\item[(i)] $\sigma_1 \in  \llbracket \varphi_2 \rrbracket_\mathscr{C}$
\item[(ii)] there exists a sequence of simplicial complexes, $\sigma_1^0, \sigma_1^1, \dots, \sigma_1^{n+1}$ with $\sigma_1^i \mathscr{C} \sigma_1^{i+1}$  for all $0 \leq i \leq n$ such that $\sigma_1 = \sigma_1^0$ and $\forall i \leq n: \sigma_1^i \in \llbracket \varphi_1 \rrbracket_\mathscr{C} $ and $\sigma_1^{n+1} \in \llbracket \varphi_2 \rrbracket_\mathscr{C} $.

\end{enumerate}
In case $(i)$, by inductive hypothesis, we have $\sigma_2 \! \in \! \llbracket \varphi_2 \rrbracket_\mathscr{C} $; hence,  $\sigma_2 \! \in \! \llbracket \varphi \rrbracket_\mathscr{C} $. \\
In case $(ii)$, by repeatedly applying the property $b$.\ of branching bisimulation equivalence, we can construct a matching sequence from $\sigma_2$.  %
The simplest case is when the set contains only $\sigma_2$ and $ \sigma_1^{n+1}  \leftrightarroweq_{br}^{\mathscr{C}} \sigma_2$. In this case, the matching sequence consists just of $\sigma_2$ and $\sigma_2 \in \llbracket \varphi \rrbracket_\mathscr{C} $ follows by induction.
Otherwise, there exists a sequence of simplices $\sigma_2^0, \sigma_2^1, \dots \sigma_2^{m+1} $ with $ \sigma_2^i \mathscr{C} \sigma_2^{i+1}$, for all $0 \leq i \leq m$, and $\sigma_2 = \sigma_2^0$ and by the stuttering lemma (Lemma~\ref{LemmaStuttering}) $\sigma_1 \leftrightarroweq_{br}^{\mathscr{C}} \sigma_2^i$ for all $i \leq m$ and $\sigma_1^{n+1}  \leftrightarroweq_{br}^{\mathscr{C}} \sigma_2^{m+1}$. From the inductive hypothesis, we have that $\sigma_2^i \in \llbracket \varphi_1 \rrbracket_\mathscr{C}$ for all $i \leq m$,  and $\sigma_2^{m+1} \in \llbracket \varphi_2 \rrbracket_\mathscr{C} $. From this, $\sigma_2 \in \llbracket \varphi \rrbracket_\mathscr{C} $ follows.

\end{enumerate}

\noindent
($``\Leftarrow"$) We prove that $ \sim^{\mathscr{C}}_{L_R} $ is a branching bisimulation. Let us consider $\sigma_1, \sigma_2\in \mathcal{K}$ such that  $\sigma_1 \sim^{\mathscr{C}}_{L_R} \sigma_2$. Since the relation is symmetric,  we have to prove that:
\begin{enumerate}
    \item[$(i)$] $\sigma_1\equiv_{\nu} \sigma_2$;
    \item[$(ii)$] for any $\sigma_1'$ such that $\sigma_1\mathscr{C} \sigma_1'$, either $\sigma_1' \sim^{\mathscr{C}}_{L_R} \sigma_2$ or there exists $\sigma_2''$ and $\sigma_2''$ such that $\sigma_2\mathscr{C}^{*} \sigma_2''\mathscr{C} \sigma_2'$, $\sigma_1\sim^{\mathscr{C}}_{L_R} \sigma_2''$ and $\sigma_1'\sim^{\mathscr{C}}_{L_R} \sigma_2'$.
\end{enumerate}

\noindent
We can observe that $(i)$ follows directly from the definition of $\sim^{\mathscr{C}}_{L_R}$. To prove $(ii)$ let us consider a $\sigma_1'$ such that $\sigma_1 \mathscr{C} \sigma_1'$. We can assume that $\sigma_1'\not \sim^{\mathscr{C}}_{L_R}\sigma_2$, otherwise the property is trivially satisfied.
We let $\mathbf{Q}$ be the set of distinct sequences (without loops) simplicial  complexes $\tilde{\sigma}=\hat{\sigma}^0,\dots, \hat{\sigma}^n$ ($n>0$) such that $\sigma_2=\hat{\sigma}^0$ and for any $i<n$ $\hat{\sigma}^0\equiv_{\nu} \hat{\sigma}^i$, $\hat{\sigma}^i\mathscr{C} \hat{\sigma}^{i+1}$ and $\sigma_1'\equiv_{\nu} \hat{\sigma}^n$.
First of all, we can observe that, since $\sigma_1 \sim^{\mathscr{C}}_{L_R}\sigma_2$, $\mathbf{Q}$  cannot be empty. In the case, a formula of the form $\top~\mathcal{R}~\varphi$ could exist that is satisfied by $\sigma_1$ and it is not satisfied by $\sigma_2$.
We have that, due to the fact that $\mathcal{M}$ is with bounded reachability, $\mathbf{Q}$ is also finite. Indeed, in $\mathbf{Q}$ can only occur simplicial complexes that satisfies the same atomic proposition of either $\sigma_1'$ or $\sigma_1$.
Let us now assume that for any $\sigma_2'$ and $\sigma_2''$ such that $\sigma_1\mathscr{C}^{*}\sigma_2''\mathscr{C} \sigma_2'$ either $\sigma_1\not \sim^{\mathscr{C}}_{L_R}\sigma_2''$ or $\sigma_1'\not \sim^{\mathscr{C}}_{L_R}\sigma_2'$.
We can split $\mathbf{Q}$ in two sets: $\mathbf{Q}_s$ and $\mathbf{Q}_f$.
The former contains the sequences where at least one internal state is not $\sim^{\mathscr{C}}_{L_R}$-equivalent to $\sigma_1$,  while the latter contains the sequences terminating in a state that is not $\sim^{\mathscr{C}}_{L_R}$-equivalent to $\sigma_1'$.
Since  $\mathbf{Q}$ is finite, both $\mathbf{Q}_s$ and $\mathbf{Q}_f$ are finite.
We can observe that, for each sequence $\tilde{\sigma}\in \mathbf{Q}_s$ there exists a formula $\varphi^s_{\tilde{\sigma}}$ that is satisfied by $\sigma_1$ but not by all the non-final simplicial complexes in $\varphi_{\tilde{\sigma}}$.
We let $\varphi_s$ be the conjunction of all $\varphi^s_{\tilde{\sigma}}$ for any $\tilde{\sigma}\in \mathbf{Q}_s$.
Similarly, for each sequence $\tilde{\sigma}\in \mathbf{Q}_f$ there exists a formula $\varphi^f_{\tilde{\sigma}}$ that is satisfied by $\sigma_1'$ but not by the final simplicial complex of $\varphi_{\tilde{\sigma}}$.
We let $\varphi_f$ be the conjunction of all $\varphi^f_{\tilde{\sigma}}$ for any $\tilde{\sigma}\in \mathbf{Q}_f$.
It is easy to see that can observe that $\sigma_1\in\llbracket \varphi_s~\mathcal{R}~\varphi_f\rrbracket$ while $\sigma_2\not\in\llbracket\varphi_s~\mathcal{R}~ \varphi_f\rrbracket$.
This contradicts the assumption that $\sigma_1 \sim^{\mathscr{C}}_{L_R} \sigma_2$. Hence, there exists at least a sequence in $\mathbf{Q}$ such that each internal state is $\sim^{\mathscr{C}}_{L_R}$-equivalent to $\sigma_1$ while the final state is $\sim^{\mathscr{C}}_{L_R}$-equivalent to $\sigma_1'$. Namely, there exist $\sigma_2''$ and $\sigma_1$ such that $\sigma_2\mathscr{C}^{*} \sigma_2''\mathscr{C} \sigma_2'$, $\sigma_1\sim^{\mathscr{C}}_{L_R} \sigma_2''$ and
$\sigma_1' \sim^{\mathscr{C}}_{L_R}\sigma_2'$.
This means that $\sim^{\mathscr{C}}_{L_R}$ is a branching bisimulation.
\end{proof}

\section{Related Work}%
\label{sec:RelatedWork}

In this section, we present an overview of the existing logic dealing with spatial aspects of systems. The origins of spatial logic can be traced back to the previous century when McKinsey and Tarski recognised the possibility of reasoning on space using topology as a mathematical framework for the interpretation of modal logic.
Formulas are interpreted in the powerset algebra of a topological space. For a thorough introduction, we refer to  ``Discrete spatial models" chapter of Handbook of Spatial Logics~\cite{smyth2007discrete}.
To develop a formalism capable of feasible model checking on such topological spaces, recent developments have led to polyhedral semantics for modal logic~\cite{bezhanishvili2018tarski,gabelaia2018axiomatization,bezhanishvili2021geometric}.

In the literature, spatial logics typically describe situations in which modal operators are interpreted syntactically against the structure of agents in a process calculus like in the Ambient Calculus~\cite{cardelli2000anytime}, whose used spatial structures are ordered edge labeled trees or concurrent systems~\cite{caires2003spatial}.

Other spatial logics have been developed for networks of processes~\cite{reif1985multiprocess}, rewrite theories~\cite{bae2012rewriting},~\cite{meseguer2008temporal}, and data structure as graphs~\cite{cardelli2002spatial}, bigraphs~\cite{conforti2007static}, and heaps~\cite{brochenin2012almighty}. Logics for graphs have been studied in the context of databases and process calculi such as~\cite{cardelli2002spatial,gadducci2007graphical}, and for collective adaptive systems~\cite{de2015logic}.
In~\cite{galton1999mereotopology,galton2003generalized,galton2014discrete}, digital images are studied by using models inspired by topological spaces without any generalising and specialising these structures, while~\cite{randell2012discrete} presents discrete mereotopology, i.e., a first-ordered spatial logic that fuses together the theory of parthood relations and topology to model discrete in image-processing applications. These results inspired works
 that use Closure Spaces, a generalisation of topological spaces, as underlying models for discrete spatial logic~\cite{ciancia2014specifying,massink2017model}.
These approaches resulted in the definition of the Spatial Logic for Closure Spaces (SLCS), and a related model checking algorithm. Even more challenging is the combination of spatial and temporal operators~\cite{kontchakov2007spatial} and few works exist with a practical perspective. Among those, SLCS has been extended with a branching time temporal logic in STLCS~\cite{ciancia2016spatial,ciancia2015experimental} leading to an implementation of a spatio-temporal model checker. Furthermore, to express in a concise way complex spatio-temporal requirement, Bartocci \textit{et al.} introduced the Spatio-Temporal Reach and Escape Logic (STREL), a formal specification language~\cite{bartocci2017monitoring}. All these logical frameworks are mainly based on graphs and suffer of the limitation already described in the introduction.
The idea of using model checking, and in particular spatial or spatio-temporal model checking is relatively recent. However, such approach has been applied in a variety of domains, ranging from Collective Adaptive Systems~\cite{ciancia2014data,ciancia2015exploring, ciancia2016tool} to signals~\cite{nenzi2017qualitative} and medical images~\cite{buonamici2019spatial,belmonte2017topological}.
All the above mentioned approaches are focussed either on a representation of the space via terms of a process algebras, or in terms of \emph{graphs}. This somehow limits the kinds of spatial relations that can be modelled. To overcome this problem, we proposed models based on \emph{simplicial complexes}.

Also bisimulation relations for spatial models are not new.  For instance, in~\cite{van2007modal} a topological bisimulation has been proposed that can be applied to the topological Kripke frames~\cite{davoren2007topological}. In this work, we consider a specific topological space, the simplicial complexes, with the goal to study the expressiveness of the proposed logic. A result similar to ours is presented in~\cite{linker2020analysing}, where a sound, but not complete, characterisation of SLCS~\cite{massink2017model} is presented.

\section{Conclusion and Future Work}%
\label{Conclusion}

The global behaviour of a system, the result of interactions among its components, is strictly related to the spatial distribution of entities. Therefore, the characterisation and verification of spatial properties play a fundamental role. In this paper, to verify the properties of surfaces and volumes or properties of systems regardless of the number of entities involved, we have defined a spatial logic on simplicial complexes. Following up on the research line of Ciancia \textit{et al.}~\cite{massink2017model}, we have introduced two logic operators, neighborhood, $\mathcal{N}$, and reachability,  $\mathcal{R}$. Intuitively, a simplex satisfies $\mathcal{N}\phi$ if it is adjacent to another simplex satisfying the property $\phi$, while a simplex satisfies $\varphi_1\mathcal{R} \varphi_2$ if it satisfies the property $\varphi_2$ or it can ``reach" a simplex that satisfies the property $\varphi_2$ by a chain of adjacent simplices satisfying $\varphi_1$.  We have defined correct and complete model checking procedures, which are linear to the dimension of the simplicial complex and the logical formula. Moreover, we have extended the concepts of bisimulation and branching bisimulation over simplicial complexes to characterise our spatial logic in terms of expressivity. We have proved that the standard boolean operators equipped with the neighbourhood operator are equivalent to the (strong) bisimulation. Instead, the standard boolean operators with the neighbourhood operators are equivalent to the branching bisimulation.
As an immediate continuation of this work, we intend to apply our spatial logic to real cases and investigate theoretical applications. In the engineering phase of the cyber-physical systems, the logic can verify if the system satisfies some constraints, such as particular constraints that involve relationships among entities. Moreover, it can be useful in understanding which spatial configurations promote the interaction between two biomolecules. Such a result is the first step towards discovering the mechanism in tumour cells.
From a theoretical application point of view, we plan to use our logic to formalise standard algebraic topology concepts over simplicial complexes, such as Betti Numbers~\cite{munkres2018elements}. These are numbers associated to simplicial complexes in terms of a \emph{topological property}, namely the number of $k$-dimensional holes. For instance, given a simplicail complex, its Betti number $0$ is the number of disconnected components; its Betti number $1$ is the number of loops; while its Betti number $2$ is the number of voids; and so far.
These numbers are largely used to identify patterns in a topological space.
Our goal is to render Betti numbers in terms of a logical formula in our framework.
This will permit identifying complexes of a given number via spatial model checking.

Another important direction is to consider temporal reasoning with spatial verification to address system evolution and dynamics within a single logic defining a spatial-temporal logic. Therefore, we will investigate theoretical aspects and the efficiency of model checking algorithms. A further promising direction is to define operators to reduce the complexity of a simplicial model preserving the bisimulation and branching bisimulations.

Finally, we plan to use standard and well-known algorithms already defined to check \emph{bisimulation}~\cite{KS90} and \emph{branching bisimulation}~\cite{groote2016m} in the context of transitions systems to check \textit{$\mathscr{C}$ bisimulation} and \textit{$\mathscr{C}$-branching bisimulation}, respectively. These adaptations will be useful to reduce the size of large-scaled systems and to discover different kind of \emph{spatial symmetries} in a spatial model.

\section*{Acknowledgment}
This research has been partially supported by the Italian PRIN project ``IT-MaTTerS'' n, 2017FTXR7S, and by POR MARCHE FESR 2014-2020, project ``MIRACLE'', CUP B28I19000330007.



\bibliographystyle{alphaurl}

\bibliography{Spatial-logic}

\end{document}